\documentclass[a4,12pt]{article}

\usepackage{float}
\usepackage{hyperref}
\usepackage{mathptmx}       % selects Times Roman as basic font
\usepackage{helvet}         % selects Helvetica as sans-serif font
\usepackage{courier}        % selects Courier as typewriter font
\usepackage{makeidx}         % allows index generation

\usepackage{amscd,amssymb,amsmath,latexsym,bm}
\usepackage[mathcal,mathscr]{euscript}
\usepackage{lipsum}
\usepackage{amsfonts}
\usepackage{graphicx}
\usepackage{epstopdf}
\usepackage{algorithmic}
\usepackage{hyperref}
\usepackage{xcolor}
\usepackage{enumerate}
\usepackage{ulem}
\usepackage{gensymb}			%DJA added this for degrees symbol

\newtheorem{theorem}{Theorem}[section]
\newtheorem{lemma}[theorem]{Lemma}
\newtheorem{corollary}[theorem]{Corollary}
\newtheorem{proposition}[theorem]{Proposition}

\newtheorem{definition}[theorem]{Definition}

\numberwithin{equation}{section}
\numberwithin{figure}{section}

\newcommand{\CM}{{\mathbb C}}
\newcommand{\NM}{{\mathbb N}}
\newcommand{\RM}{{\mathbb R}}

\newcommand{\ZM}{{\mathbb Z}}

\newcommand{\Aa}{{\mathcal A}}

\newcommand{\Bb}{{\mathcal B}}

\newcommand{\Ww}{{\mathcal W}}

\newcommand{\Ss}{{\mathcal S}}

\newcommand{\Tr}{\mbox{\rm Tr}}
\newcommand{\Tt}{{\mathcal T}}

\newcommand{\Kk}{{\mathcal K}}

\newcommand{\tr}{{\rm tr}}

\newcommand{\spec}{{\rm spec}}

\begin{document}

\title{Bulk-boundary correspondance for Sturmian Kohmoto like models}

\author{Johannes Kellendonk \\
\texttt{\small Universit\'e de Lyon,
Universit\'e Claude Bernard Lyon 1} \\
\texttt{\small and
Institute Camille Jordan,} \\
\texttt{\small CNRS UMR 5208, 69622 Villeurbanne, France}
\and
Emil Prodan \\
\texttt{\small Department of Physics and 
Department of Mathematical Sciences,} \\
\texttt{\small Yeshiva University, New York, USA}}

\maketitle

\begin{abstract}
We consider one dimensional tight binding models on $\ell^2(\ZM)$ 
whose spatial structure is encoded by a Sturmian sequence 
$(\xi_n)_n\in \{a,b\}^\ZM$. An example is the Kohmoto Hamiltonian, which is given by the discrete Laplacian plus an onsite potential 
$v_n$ taking value $0$ or $1$ according to whether $\xi_n$ is $a$ or $b$. The only non-trivial topological invariants of such a model are its gap-labels. The bulk-boundary correspondence we establish here states that there is a correspondence between the gap label and a winding number associated to the edge states, which arises if the system is augmented and compressed onto half space $\ell^2(\NM)$. 
%Indeed, depending on a circle-valued parameter the energies of the edge states wind around a gap and so define a winding number. 
This has been experimentally observed with polaritonic waveguides. A correct theoretical explanation requires, however, first a smoothing out
of the atomic motion via phason flips. With such an interpretation at hand, the winding number corresponds to the mechanical work through a cycle which the atomic motion exhibits on the edge states. 
\end{abstract}

%%\pacs{73.43.Cd,73.43.Nq,74.62.En,75.85.+t}

%\tableofcontents

\section{Introduction}

Inspired by the research on topological insulators \cite{Hal,KM1,KM2,BHZ,KWB,MB,FK,HQW}, there is growing experimental effort to search for topological boundary resonances in many kinds of wave-supporting media, such as photonic \cite{RZP,WCJ,HMF} and phononic \cite{PP,KL,PCV,NKR} crystals, as well as plasmonic \cite{SR} systems. A more recent and quite vigorous effort of the community is to go beyond the periodic table of topological insulators and superconductors \cite{SRFL,RSFL,Kit}, and try to detect such topological boundary resonances in aperiodic systems, such as almost-periodic \cite{KLR,MBB,KRZ,Pro,HPW}, quasi-crystalline \cite{KZ,VZK,TGB,TDG,VZL,LBF,DLA,BRS,BLL} and amorphous patterns \cite{MNH}. This is now experimentally feasible because, with the new wave-supporting media mentioned above, one has nearly perfect control over the design of the system, something which is not at all the case for the electronic systems.

\vspace{0.2cm}

Quite generally, the bulk-boundary correspondence principle relates the topological invariants of the bulk to topological invariants of the boundary of a physical system. Since the boundary invariants are carried by boundary states, this relation explains the emergence of robust boundary spectrum in systems with non-trivial bulk-boundary correspondence. This has long been understood for the quantum Hall effect \cite{Hat,SKR,KRS,KS} and, while the bulk-boundary principle is relatively well understood for the almost-periodic cases \cite{Pro}, it has never been rigorously formulated for the quasi-crystalline and amorphous cases. Here we deal exclusively with the one dimensional quasi-crystalline case and, as we shall see, the bulk-boundary correspondence is quite subtle. 
One of the basic results about the topology of quasi-crystals is that the phason degree of freedom lives in a Cantor set \cite{Le,FHKphys}, which is a totally disconnected space. In contradistinction, for almost periodic systems, the phason lives on a circle. While that Cantor set differs from the circle only on a set of measure zero, this difference fundamentally changes the topology of both the bulk and boundary for the quasi-crystalline systems. Indeed, while the bulk gap labels remain unchanged, they no longer can be interpreted in terms of Chern numbers as done for almost periodic systems \cite{KLR,Pro}. Furthermore, the topological structure of the boundary states is quite different, in particular, any unitary operator constructed from the boundary states is stably homotopic to the identity. Hence, we are facing a situation where the bulk-boundary principle is trivial in the sense that all topological bulk invariants, provided by the bulk gap labels, are sent to the unique trivial boundary invariant. This is quite unfortunate because the one dimensional quasi-crystalline systems are particularly interesting as their bulk energy spectrum is expected to be a Cantor set of zero measure, hence the spectral gaps abound and, as we shall see, all existing gaps have non-trivial labels. 

\vspace{0.2cm}

Recent experiments \cite{TGB,BLL} with polaritonic wave guides structured like Fibonacci chains have, nevertheless, reported interesting findings. In \cite{TGB}, the resonance spectrum of the cavity was measured and its bulk gaps were labeled in agreement with the gap-labeling theorem \cite{Bel86,Bel95}. Furthermore, in \cite{BLL}, by coupling the cavity with its mirror image, localized modes were created at the center of the mirror symmetry and their frequencies were plotted against the phason. From these plots, a boundary winding number was read off and shown to coincide with the bulk gap labels. However, as we shall see, the set of boundary eigenvalues for all different values of the phason has zero measure and, as a consequence, the boundary spectrum always displays gaps. As such, the boundary spectral patches plotted in \cite{BLL} need to be connected by imaginary lines in order to see the winding numbers. Such lines, unfortunately, are not justified by the experiments and the winding numbers depend on the choice of these lines, which is not unique at all.

\vspace{0.2cm}

The purpose of our work is to reconcile these observations with the triviality of the boundary topology and to give a mathematically rigorous definition of the winding numbers observed in \cite{BLL}, as well as to provide a rigorous explanation of the equality between the bulk gap labels and these winding numbers. The latter is based on a modification of the standard exact sequence used in the bulk-boundary correspondence, which consists, roughly speaking, in {\it augmenting} the Cantor set to a circle, an idea that goes back to Denjoy's work to classify the homeomorphisms of the circle \cite{Denjoy}. We also establish that the boundary invariant, which is here the winding number of the augmented system, is related to a physical response coefficient resulting from a change in the potential due to the phason motion,
a characteristic feature of quasi-crystals. 
This is similar but seemingly not identical to the response coefficient associated with the pressure on the boundary \cite{Kel-bf,KelZois}. Our proofs are based on algebraic topological methods. They involve the $C^*$-algebras of the bulk, half-space and boundary physical observables for the  
Kohmoto type models. The framework we use to prove the bulk-boundary correspondance is the $K$-theory of these  and related $C^*$-algebras. While this might appear to be heavy machinery, we do not know at present of any other approach, as the underlying system is aperiodic. On the other hand, the $K$- and $C^*$-algebraic approach to aperiodic solid state systems proposed by Jean Bellissard \cite{Bel86} is by now well-known and
our $K$-theory arguments are quite standard, following essentially the ideas of \cite{KRS} and \cite{KR}. 

\vspace{0.2cm}

Our paper is organized as follows. Section~\ref{sec-prelim} collects results from symbolic dynamical systems which allow one to study continuity properties of spectra of Schr\"odinger operators and underlie the non-commutative geometry of such systems. The simplest one dimensional models for quasi-crystals, the Sturmian sequences, have been studied extensively in the forties by Morse and Hedlund \cite{MH1,MH2} in this context. Section~\ref{seq-Sturmian} introduces three dynamical systems naturally associated to Sturmian sequences, namely, 1) the original Sturmian subshift $(\Xi,\ZM)$ whose space $\Xi$ is a Cantor set and whose sequences have finite local complexity, 2) the augmented subshift $(\tilde\Xi,\ZM)$ whose space $\tilde\Xi$  is topologically a circle and which contains the original subshift as its unique minimal component, and 3) an approximation system $(\Xi_\epsilon,\ZM)$ whose space $\Xi_\epsilon$ is also topologically a circle, but whose sequences have infinite local complexity. The main new results of Section~\ref{seq-Sturmian} are about the convergence (in spectrum) of the
Schr\"odinger operators associated to $(\Xi_\epsilon,\ZM)$ to those of the augmented system $(\tilde\Xi,\ZM)$ (Lemma~\ref{lem-eps}) and, in particular, the continuity of the Dirichlet eigenvalues for the augmented system in the intercept parameter (as we call the phason degree of freedom) (see Cor.~\ref{cor-cont}, which is based on the general Thm.~\ref{thm-cont}). This continuity is essential for the definition of the winding number which we give in Section~\ref{sec-wind}. Section~\ref{sec-numerics} reports numerical experiments for both full and half-space Kohmoto models, which illustrate the issues raised at the beginning of this Section and also show how our proposed solution works. More precisely, direct evidence is provided that the $K$-theoretic labels for the bulk gaps correlate with the well defined spectral flow of the boundary states of the augmented models. Section~\ref{sec-calgebras} introduces the $C^*$-algebras canonically associated with the dynamical systems introduced in Section~\ref{seq-Sturmian}, and establishes various relations between them. Section~\ref{sec-ktheory} computes the $K$-theories of these algebras and the final proof of the bulk-boundary correspondence and the physical interpretation of the boundary invariant is given in Section~\ref{sec-bulkboundary}.

\section{Preliminaries from symbolic dynamics}\label{sec-prelim}

Consider a compact metric space $X$ together with a homeomorphism $\alpha:X\to X$.
Iterating the homeomorphism defines an action of $\ZM$ on $X$, i.e.\ a group homomorphism $\ZM\ni n \mapsto \alpha^n\in Homeo(X)$. This is also referred to as a dynamical system and denoted $(X,\ZM)$. A subsystem $(Y,\ZM)$ of $(X,\ZM)$ is given by a subset $Y\subset X$ which is invariant under the action. 
The orbit of an arbitrary subset $Y\subset X$ is the subset $\{\alpha^n(y)|y\in Y, n\in\ZM\} \subseteq X$ and 
we denote by $O(Y)$ the closure of the orbit of $Y$. 
The dynamical system is called minimal if the orbit of any point $x\in X$ is a dense subset of $X$, that is, $O(x)=X$ for any $x\in X$.
Every dynamical system contains a minimal subsystem. Two minimal subsystems are either equal or do not share a common point and for this reason the minimal subsystems are also called minimal components. 

\vspace{0.2cm}

Of importance below will be the universal dynamical system $(X_u,\ZM)$ associated with $(X,\ZM)$ \cite{Bec,BBdN16}, which is constructed as follows. Let $I(X)$ be the space of closed subspaces of $X$ which are invariant under the action. We equip $I(X)$ with the Hausdorff topology, that is, two closed subsets $X_1,X_2$ of $X$ are at most Hausdorff distance $\epsilon$ apart if, within distance $\epsilon$ of each point of $X_1$ lies a point of $X_2$ and vice versa. Now the space of the universal dynamical system is
$$X_u = \{(Y,y)\in I(X)\times X \,|\, y\in Y\}$$
with subspace topology of the product topology on $I(X)\times X$. Clearly $X_u$ is compact and metrisable. $\alpha(Y, y) = (Y,\alpha(y))$ defines the $\ZM$-action on $X_u$.

\vspace{0.2cm}

Let $\Aa\subset \RM$ be a compact subset and $\Aa^\ZM$ be the space of two-sided infinite sequences with values in $\Aa$ equipped with the product topology. This topology is metrisable and we choose to work with the following metric: 
$$d(x,y) = \inf \{ \epsilon >0 \, | \ \forall \ |k|\leq \epsilon^{-1} : |x_{k}-y_k|\leq \epsilon\}$$ 
$\ZM$ acts on sequences by left shift, $\alpha(x)_n = x_{n+1}$ and this action is continuous and preserves $\Aa^\ZM$. 
%It carries over to an action on $\Dd(\Aa)$. 
$(\Aa^\ZM,\ZM)$ is a symbolic dynamical system. 
%We denote by $I(\Aa^\ZM)$ the closed shift invariant subspaces of $\Aa^\ZM$. 

\vspace{0.2cm}

%We now define the notion of strongly pattern equivariant functions. 
We denote by $\Aa^N$ the sequences of length $N$ with values in $\Aa$. In analogy to symbolic dynamics, we call these sequences also the (allowed) words of length $N$. We put on $\Aa^N$ the product topology.
\begin{definition}\label{def-blocksliding} A sliding block code of 
%length $N$ 
range $r\in\NM$ over $\Aa$ is a
continuous function $b:\Aa^{2r+1} \to \CM$. Such a function extends to a function on $\Aa^\ZM$ by reading from $x\in \Aa^\ZM$ only the block of length $2r+1$ around $0$, {\it i.e.} the word $x_{-r}\cdots x_{r}$.
A function $f:\ZM\to \CM$ is called strongly pattern equivariant for $x\in \Aa^\ZM$ if there exists a sliding block code $b:\Aa^N\to \CM$ such that $f(n) = (b\circ\alpha^{n})(x)$.
\end{definition}   
%Note that if $\Aa$ is finite then any $x\in \Aa^\ZM$ has finite local complexity and the continuity assumption on $b$ is automatic, as $b$ takes then only finitely many values. The above definition then reduces to the definition given in \cite{Kellendonk}.
%
%\vspace{0.2cm}
%
%We can turn \ref{def-blocksliding} around. 
Given a sliding block code $b$ of range $r$, we obtain, for every $x\in \Aa^\ZM$, a strongly pattern equivariant function $\tilde b_x:\ZM\to \CM$ by setting 
\begin{equation}\label{eq-b1}
\tilde b_x(n) = b(\alpha^{n}(x)) = b(x_{n-r},\cdots ,x_{n+r}).
\end{equation}
A family of Schr\"odinger operators $\{H_x\}_{x\in \Aa^\ZM}$ is called strongly pattern equivariant if there is a finite subset $S\subset\ZM^+$ and for each $k\in S$ a sliding block code $b_k$ such that
\begin{equation}\label{eq-b2} H_x = \sum_{k\in S} (\widetilde {b_k})_x T^k + h.c. 
\end{equation}
where $T$ is the left translation operator on $\ell^2(\ZM)$, $T\psi(n) =\psi(n+1)$, and $(\widetilde {b_k})_x$ act as multiplication operators on the same Hilbert space. As usual, $h.c.$ stands for the hermitian conjugate. The largest value of $S$ is the range of the operator.
A simple example is given by the model
\begin{equation}\label{eq-model}
 H_x\psi(n) =\psi(n-1) + \psi(n+1) + v_x(n) \psi(n) \end{equation}
where $v=\widetilde{b_0}$ is a strongly pattern equivariant function defined by a sliding block code $b_0:\Aa^1\to \RM$ of range $0$. The original Kohmoto model \cite{KO} is of the above form, with $\Aa = \{a,b\}$ and $b_0(a)=0$, $b_0(b)=1$.

\vspace{0.2cm}

Strongly pattern equivariant families of Schr\"odinger operators are covariant, $T H_x T^{-1} = H_{\alpha(x)}$. 
Note that there is always a uniform (in $x$) upper bound for the norm of the $H_x$. 
We can restrict such families to subsystems $(X,\ZM)\subset (\Aa^\ZM,\ZM)$ where $X$ is closed and shift invariant. We then denote by $H_X$ the restriction of $\{H_x\}_{x\in \Aa^\ZM}$ to points $x\in X$ and $\spec(H_X) = \bigcup_{x\in X} \spec(H_x)$. It is natural to view the latter as elements of $\Kk(\RM)$, the space of compact subsets of the real line equipped with the Hausdorff topology.

\vspace{0.2cm}

The map $X\ni x\mapsto H_x$ is strongly continuous. Indeed, as the distance between $x$ and $x'$ gets closer then they coincide more and more on larger pieces and hence, when applied to a compactly supported vector near the origin, $H_x$ and $H_{x'}$ yield the same result. It follows that $\spec (H_x)$ contains $\spec (H_x')$ for all $x'\in O(x)$ and hence is equal to $\spec (H_{O(x)})$. In
particular, if $(X,\ZM)$ is minimal then all $H_x$, $x\in X$ have the same spectrum. 
%The following result can be found in \cite{Bec} (see Corollary 4.3.14) for finite $\Aa$. Below we state a version that covers the case of infinite $\Aa$ in our more limited context of crossed product algebras.
%%%%%%%%%%%%%%%%%%%%%%%%%%%%%%%%%%%
\begin{theorem}[\cite{Bec,BBdN16}] \label{thm-BBdN16}
Let $\{H_x\}_{x\in \Aa^\ZM}$ be a family of strongly pattern equivariant Schr\"odinger operators.
The map $\Sigma:I(\Aa^\ZM)\to \Kk(\RM)$,
$$\Sigma(X) = \spec(H_X) $$%:= \bigcup_{x\in X} spec(H_x)$$
is continuous.
\end{theorem} 
%%%%%%%%%%%%%%%%%%%%%%%%%%%%%%%%%%%%

\noindent {\bf Proof:} For finite $\Aa$ this is Corollary~4.3.14 of \cite{Bec}. For compact
$\Aa $ the result follows with the same methods as
every sliding block code defines a continuous function on ${\Aa^\ZM}_u$ by $b(Y,y) = b(y)$.
The result is then a consequence of Thm.~2 of \cite{BBdN16}.
% (which, as we said, is $b$ evaluated on the block of length $N$ centered at $0$). This now shows that there is an element $H\in C({\Aa^\ZM}_u)\rtimes\ZM$ such that the induced representation $\pi_x$ from the evaluation representation $ev_x:C({\Aa^\ZM}_u) \to \CM$, $ ev_x(f) = f(\Aa^\ZM,x)$ satisfies $\pi_x(H) = H_x$ (as the family $H_x$ is strongly pattern equivariant, we could also take here $ev_x(f) = f(Y,x)$ for any $Y\in I(\Aa^\ZM)$ which contains $x$). Now using ideas of Blanchard, and Landsman \& Ramazan, [BBdN16] show that 
% $C({\Aa^\ZM}_u)\rtimes\ZM$ is a continuous $C^*$-field over $I(\Aa^\ZM)$, from which it follows the spectrum of $H(X)$, the evaluation of the continuous section defined by $H$ on $X\in I(\Aa^\ZM)$, is continuous of $X$. The statement then follows, because
% $$
%spec (H(X)) = \bigcup_{x\in X} spec(H_x)=spec(H_X),
%$$ 
%as the family of representations $\pi_x$ is faithful. 
\hfill q.e.d.
 \bigskip
  
We now turn to the situation with a boundary, which is formalized as follows. We add a large real number $z$ to the alphabet $\Aa$, $\hat\Aa = \Aa\cup\{z\}$ and then construct the following shift space. If $X$ is a closed invariant subspace of $\Aa^\ZM$ and $r\in \NM$, then we let $X_r^+$ be the set of sequences $x_r^+\in \hat\Aa^\ZM$ of the following form: there is $x\in X$ 
%and $k\in\ZM$ 
such that 
$$ (x_r^+)_n = \left\{\begin{array}{ll}
x_n & \mbox{if } n\geq -r,\\
z & \mbox{otherwise}.
\end{array}\right.$$
$X_r^+$ is a closed subspace of $\hat\Aa^\ZM$, but it is not shift invariant. 
%For $x\in X$ we denote by $\hat X(x)$ the closure of the orbit of $x^+$. This is now by definition a closed invariant subspace of $\hat\Aa^\ZM$. 
Note that $O(x_r^+)$ contains the constant sequence $x_n=z$ and the set $M_0$ containing only this sequence is a minimal component.
%We let $\hat X$ be the union of all the $O(x^+)$, $x\in X$; it is the closure of the orbit of $X^+$. 
If $X$ is minimal then 
$O(x_r^+) = \{\alpha^n(x_r^+)|n\in\ZM\}\cup X\cup M_0$.
This can be seen as follows: if $y\in O(x_r^+)$ lies in a minimal component then it is either an accumulation point of the forward, or of the backward orbit. But all accumulation points of the forward orbits lie in $X$ whereas the backward orbit has one single accumulation point, namely the constant sequence $x_n=z$. Lastly, let us remark that the map $x \mapsto x_r^+$ from $X$ to $X_r^+$ is  continuous and onto.

\vspace{0.2cm}

The map $p:\hat\Aa^1\to\CM$, $p(\omega) = 1$ if $\omega\neq z$ and else $0$ is a sliding block code. We use this to extend the family of strongly pattern equivariant Hamiltonians to the larger space $\hat\Aa^\ZM$,
$$H_{x,s} = \widetilde{(p\circ \alpha^{-s})}_x \left(\sum_{k\in S} (\widetilde {b_k})_x T^k\right) + (1-\widetilde{(p\circ \alpha^{-s})}_x) z + h.c. , \quad s \in \NM.$$   
Here the sliding block codes $b_k$ are extended to $\hat \Aa$ by $0$ on words which contain $z$.
%%%%%%%%%%%%%%%%%%%%%%%%%%%%%%
\begin{theorem} \label{thm-cont}
Let $(X,\ZM)$ be a minimal subsystem of $(\Aa^\ZM,\ZM)$.
The map $X_r^+ \ni x_r^+ \mapsto \spec(H_{x_r^+,s})$ is continuous. 
\end{theorem}
%%%%%%%%%%%%%%%%%%
{\bf Proof}: We show that $O(y_r^+)$ and $O({y'}_r^+)$ are close if $y_r^+$  and ${y'}_r^+$ are close in $X_r^+$. 
% For that we use the metric $d$ between sequences given by the inverse of the coincidence radius $R(x,x')= \max\{n\in \NM\,|\, \forall |k|\leq n: x_k=x'_k\}$.
Let $x\in O(y_r^+)$. We have $d(x,O({y'}_r^+)) = \inf \{d(x,x')|x'\in O({y'}_r^+)\}$. If $x$ is in a minimal component of $O({y}_r^+)$ then $d(x,O({y'}_r^+)) = 0$, because $O({y}_r^+)$ and $O({y'}_r^+)$  have the same minimal components. So let $x$ not be in a minimal component, and hence $x = \alpha^n(y_r^+)$ for some $n \in \ZM$. 

\vspace{0.2cm}

Suppose that $d(y_r^+,{y'}_r^+)<\epsilon$, that is, $|(y_r^+)_k-({y'}_r^+)_k|< \epsilon$ for all $|k|<\epsilon^{-1}$. If $0\leq n \leq \frac12 \epsilon^{-1}$ then $d(\alpha^n(y^+),\alpha^n({y'}_r^+))<2\epsilon$. If $ n \geq \frac12 \epsilon^{-1}$ then $d(\alpha^n(y_r^+),\alpha^n({y}))<2\epsilon$ and since $y\in X\subset O({y'}_r^+)$ we see that, in both cases, $d(\alpha^n(y_r^+),O({y'}_r^+))<2\epsilon$. Finally, if $n$ is negative then $d(\alpha^{-n}(y_r^+),\alpha^{-n}({y'}_r^+)) \leq d(y_r^+,{y'}_r^+)$. This shows that the Hausdorff distance between $O({y}_r^+)$ and $O({y'}_r^+)$ is bound by twice the distance between $y_r^+$ and ${y'}_r^+$.

\vspace{0.2cm}

By Theorem~\ref{thm-BBdN16} the map $\Sigma: I(\hat \Aa^\ZM)\to \Kk(\RM)$ is continuous. We apply this to the subspace $\{O(x_r^+)|x_r^+\in X_r^+\}\subset I( \hat \Aa^\ZM)$ to conclude with the above that 
$x_r^+\mapsto \spec (H_{O(x_r^+),s})$ is continuous. We saw that  $\spec (H_{O(x_r^+),s})= \spec(H_{x_r^+,s})$. \hfill q.e.d. 
\bigskip
%
%We remark that the proof of the theorem extend to the situation, in which $X$ has not necessarily only one minimal component, but each $X(y)$ has the same minimal components and thus contains besides these only the orbit of $y^+$.
%\bigskip

\vspace{0.2cm}

For $x\in X$ we define $\hat H_x$ to be the compression of $H_x$ to the right half-space, i.e.\
$$ \hat H_x :\ell^2(\NM) \rightarrow \ell^2(\NM), \quad \hat H_x = \Pi^\ast H_x \Pi,$$
with $\Pi$ the usual partial isometry from $\ell^2(\NM)$ to $\ell^2(\ZM)$. In the case of strongly pattern equivariant Schr\"odinger operators 
$\hat H_x$ is the restriction of  
$\left(\sum_{k\in S} (\widetilde {b_k})_x T^k\right) +  h.c.$ to $\ell^2(\NM)$.
%%%%%%%%%%%%%%%%%%%%%%%%%
\begin{corollary}\label{cor-cont1}
Let $(X,\ZM)$ be a minimal subsystem of $(\Aa^\ZM,\ZM)$ and $\{H_x\}_{x \in X}$  a strongly pattern equivariant family of Schrödinger operators.
Then the map $X \ni x \mapsto \spec(\hat H_x)$ is continuous. 
\end{corollary} 
%%%%%%%%%%%%%%%%%%%%%%%%
{\bf Proof:} Recall that the map $x \mapsto x_r^+$ is continuous and take $R \in \NM$ greater than the maximum range of sliding blocks in $H_x$. Then $H_{x_R^+,R} = \hat H_x \oplus z(I - \Pi \Pi^\ast)$ hence $\spec(H_{x_R^+}) = \spec(\hat H_x)\cup\{z\}$. If we take $z$ larger than a common bound on the norm of the $\hat H_x$, then its contribution to the spectrum is well separated from $\spec(\hat H_x)$. The latter is then continuous of $x$, which follows from the continuity of $\spec(H_{x_R^+,R})$ w.r.t. $x_R^+$.\hfill q.e.d.

%%%%%%%%%%%%%%%%%%%%%%%%%%%%%%%%%%%%%

\section{Dynamical systems and models associated to Sturmian sequences}\label{seq-Sturmian}
\begin{figure}
\includegraphics[width=1\textwidth]{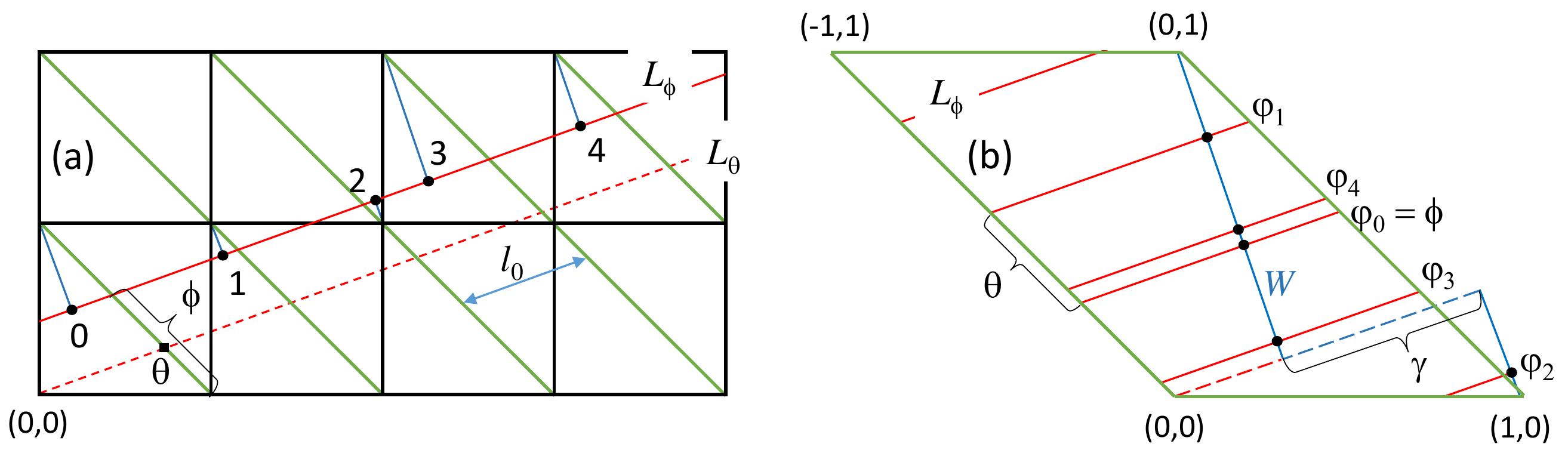}
\caption{{\small The cut \& project scheme. (a) A one dimensional quasi-periodic point pattern (the dots $0,1,2,\ldots$) is generated by projecting the points of 
$\ZM^2 \cap \big ([0,1]^2+L_\phi \big )$ orthogonally onto the line $L_\phi$.
(b) When wrapped on $\RM^2/\ZM^2$, the same point pattern is generated by the intersection of  $L_\phi$ (which now winds on the torus) with the transversal $W$ shown as a blue line (dashed part excluded). The figure also illustrates the notation introduced and used in the text.}}
\label{Fig-Sturmian}
\end{figure}

Sturmian sequences can be defined as the repetitive aperiodic sequences $x\in \{a,b\}^\ZM$ which have minimal word complexity function, namely there are exactly $n+1$ distinct words of length $n$ in $x$. In the seminal work by Morse and Hedlund \cite{MH1,MH2}, various other characterizations of these sequences have been given and here we make substantial use of their geometric characterization by means of the cut \& project formalism, illustrated in Fig.~\ref{Fig-Sturmian}. Consider the integer lattice $\ZM^2$ in $\RM^2$ and parametrize the anti-diagonal $D$ of the unit cell $[0,1]^2\subset\RM^2$ by $[0,1]$: 
$[0,1]\ni \varphi \mapsto i(\varphi):=\varphi e_2 + (1-\varphi) e_1\in D$.
Let $\theta\in [0,1]$ be an irrational and 
$L_\theta$ be the line in $\RM^2$ which passes through the origin $0\in\RM^2$ and the point on $D$ parametrized by $\theta$. Consider the orthogonal projection $\pi^\perp$ onto the orthocomplement of $L_\theta$ and let $W$ be $-\pi^\perp([0,1]^2)$. These are the data of the cut \& project scheme.
To obtain a Sturmian sequence with intercept $\phi\in [0,1]$ consider the intersection of $L_\phi:= L_\theta+i(\phi-\theta)$ with $\ZM^2+W$. This is a discrete subset of $L_\phi$ 
which can be ordered, as $L_\phi$ is a line, and so is given by a sequence $w(\phi)=\big (w_n(\phi)\big )_{n \in \ZM}$ of points. Each of these points can be uniquely specified by their $\varphi_n$ coordinate on $D$.
% We always fix the origin of $L_\phi$ and the labeling such that $w_0=0$ and $\varphi_0 = \phi$. 
Denote by $NS$ the set of $\phi$ for which $L_\phi$ does not intersect a boundary point of $\ZM^2+W$. 
It turns out that, for all $\phi\in NS$, the difference vectors $w_n-w_{n-1}$ take only two different values \cite{MH1,MH2}; we can distinguish them by their length which we denote $a$ and $b$ with $a<b$. Defining $\xi_n$ to be the length of $w_n-w_{n-1}$, we obtain in this case a sequence $\xi=(\xi_n)_n\in \{a,b\}^\ZM$ which has the desired minimal word complexity. 
To note its dependence on the intercept we also write $\xi(\phi)$.

%                               Figure Sturmian sequence

\vspace{0.2cm}

For $\phi\in NS$ and with the notation from Fig.~\ref{Fig-Sturmian}, we can give the formula:
\begin{equation}\label{Eq-SturmPattern}
w_n(\phi)=n \, l_0 + \gamma \big(\varphi_n - \chi(\varphi_n-\theta) \big ) , \ \  \varphi_n = \big\{ \phi + n \theta \big \} , \quad n\in \ZM .
\end{equation}
Here $\{r\}$ denotes the fractional part of $r\in \RM$, $\chi$ the indicator function on $\RM^+$. Note that the function $r\mapsto \{r\}-\chi(\{r\}-\theta)$ is $0$ on $r=0$ and on $r=1$ and has a discontinuity at $r=\theta$. When $\theta = \frac{3-\sqrt{5}}{2}$, $b/a$ becomes the golden ratio and $w(\phi)$ describes a Fibonacci sequence.

\begin{figure}
\center
\includegraphics[width=0.5\textwidth]{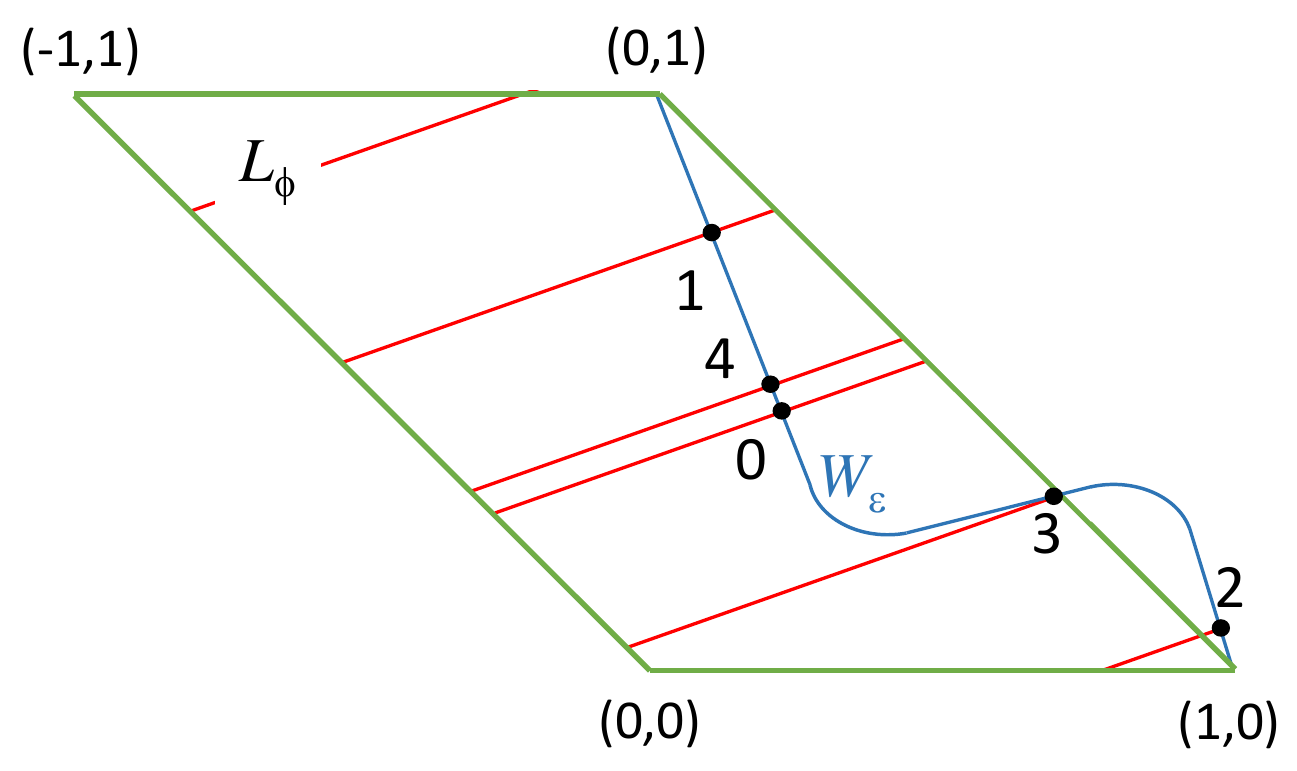}
\caption{{\small Example of a smoothed Sturmian system.}}
\label{Fig-SmoothSturmian}
\end{figure}

The Sturmian sequences obtained in the above geometric way, for fixed $\theta$ but varying intercept $\phi$ (but such that $L_\phi$ does not intersect the boundary of $\ZM^2+W$)
form a shift invariant non-closed subset of $\{a,b\}^\ZM$. 
%This minimal component $(\Xi,\ZM)$ is called the Sturmian subshift with parameter $\theta$. 
From a topological point of view, it is most interesting to study what has to be added to this system to get a complete dynamical system. There are several options.

\begin{enumerate}
\item 
Identifying the end points of $[0,1]$ to obtain a circle, $NS$ becomes a dense set of $S^1$ and we refer to its points as the non-singular values for $\phi$. One easily sees that its complement of so-called singular values is $\theta\ZM$ modulo $1$.
One way to obtain a completion of the dynamical system $(NS,\ZM)$ is to complete $NS$ in the euclidean topology to obtain all of $S^1$. 
The Sturmian sequence with intercept $\phi$ shifted by one unit (to the left) corresponds to the sequence with intercept $\phi+\theta$.
The completed system is therefore the rotation action by $\theta$ on the circle $S^1$. It is minimal as we assume $\theta$ to be irrational.
%This will play a minor role but is for reference and to understand what is wrong in the literature.
\item From the point of view of symbolic dynamics it is natural to complete the shift invariant subset of sequences $\xi(\phi)$ with intercept $\phi\in NS$ in the topology of $\{a,b\}^\ZM$.
The result is, by definition, the Sturmian subshift with parameter $\theta$. We will denote it by 
$(\Xi,\ZM)$.
It is a minimal system.
But its space $\Xi$ is totally disconnected, as $\{a,b\}^\ZM$ is a Cantor set. We can describe $\Xi$ in a geometric way by means of limits. Indeed, let $\phi$ be a singular value. Then we can approach it from the left or from the right by sequences of non-singular values. While on $S^1$ the limits are the same, $\xi(\phi^+):=\lim_{\phi\to\phi^+}\xi(\phi)$ will differ from $\xi(\phi^-):=\lim_{\phi\to\phi^-}\xi(\phi)$ by exactly one word of length $2$: the first can be obtained from the second by flipping exactly one pair $ab$ to $ba$. We call such Sturmian sequences also singular; they are characterised by the fact that they contain a flipping pair, that is a pair ($ba$ or $ab$) which can be flipped so that the result remains a Sturmian sequence. 

Geometrically we can think of $\Xi$ as the cut up circle: Whenever $\phi$ is singular we replace it by its two half-sided limits $\phi^+$ and $\phi^-$. This can be done topologically by means of inverse limits: Consider a chain of finite subsets $S_k\subset S_{k+1}$ such that $\bigcup_{k\in\NM} S_k $ consists of all singular points. At step $k$ take out the points of $S_k\backslash S_{k-1}$ and complete the individual components by adding boundary points. In the limit we have disconnected the circle along $\theta\ZM$. 

\item 
If, instead of replacing a singular point $\phi$ by its two half-sided limits $\phi^+$ and $\phi^-$ we replace it by a closed interval $[0,1]$ in the above construction, we obtain the augmented system $(\tilde\Xi,\ZM)$. It can be parametrized as follows: It contains $NS$ and for each singular $\phi$ we parametrize the added in interval by $(\phi,t)$, $t\in[0,1]$. $\tilde\Xi$ contains $\Xi$ as a closed subspace where, for each singular $\phi$, $\phi^- = (\phi,0)$ and $\phi^+ = (\phi,1)$.

Let us give this a symbolic interpretation. If we would construct a sequence of points $(w_n)_n$ as above for a singular choice of the intercept, then there would be exactly one position $n$ at which $w_n-w_{n-1}$ has a third length $c=b-a$, and 
both, $w_{n+1}-w_{n}$ and $w_{n-1}-w_{n-2}$ would have length $a$. Therefore the sequence $(w_n-w_{n-1})_n$ would not be a Sturmian sequence. However, if we take out either $w_n$ or $w_{n-1}$ then the length of the difference vectors form a Sturmian sequence. 
(The above describe limiting procedure can be interpreted in making this choice.)
The augmented Sturmian sequence can now be symbolically understood by saying that instead of taking out one of the two points one replaces the two by a convex combination $tw_n+(1-t)w_{n-1}$, $t\in [0,1]$. $(\phi,t)$ therefore corresponds to a sequence $\xi(\phi,t) \in [a,b]^\ZM$ in which the flipping pair $ba$ has been replaced by the word $b_t a_t$ where $b_t = ta + (1-t)b$ and $a_t = (1-t)a + tb$.

$(\Xi,\ZM)$ is the unique minimal component of $(\tilde\Xi,\ZM)$. 
The augmented version is crucial for a correct understanding of the bulk-boundary correspondance. 
\item For numerical approximation and comparison with other results from the literature we also consider a smoothed out version of the Sturmian subshift. To describe this it is convenient to consider the cut \& project scheme on the torus $\RM^2/\ZM^2$, as in Fig.~\ref{Fig-Sturmian}(b). There, the line $L_\phi$ wraps densely around this torus and $W$ appears as a one-dimensional line segment of constant irrational slope. The constant slope guaranties that the resulting sequence $(w_n-w_{n-1})_n$ has only finitely many values, but inhibits (because of the irrationality of the slope) that the segment is a closed continuous curve on the torus. On the other hand, if $W$ is deformed into a closed continuous curve, which is transversal to $L_\phi$ for all $\phi$, then the map $\phi\mapsto \xi(\phi)$ from above becomes continuous for all $\phi\in S^1$. As a consequence, for a transversal $W$ which is a closed continuous curve the corresponding symbolic dynamical system is topologically conjugate to the rotation by $\theta$ on $S^1$. One should be aware, however, that $\xi(\phi)$ takes now a dense set of values in the interval $[a,b]$, hence the system has infinite local complexity \cite{KellendonkLenz}.

We consider here a deformation $W_\epsilon$ of $W$ which is obtained if one replaces the 
indicator function $\chi$ in (\ref{Eq-SturmPattern}) by 
%a continuous function $\chi_\epsilon$ which satisfies $|\chi(t)- \chi_\epsilon(t)|\leq \epsilon$ pour tout $t\in (-\epsilon,\epsilon)$. For numerical purposes we will take the function
%Heavyside step function 
$\chi_\epsilon(t)=\frac12(1+\tanh(\frac{t}{\epsilon}))$. 
Indeed, if we plot the segment $W$ as a function of $\phi$ with an $y$-axis which is along $L_\theta$ (thus a bit tilted) then it has a discontinuity at $\phi=\theta$. We smooth this discontinuity out with the help of the above approximation to the step function (see Fig.~\ref{Fig-SmoothSturmian}). 
We denote the corresponding system by  $(\Xi_\epsilon,\ZM)$ and the sequence with intercept $\phi$ by $\xi(\epsilon,\phi)$. $(\Xi_\epsilon,\ZM)$ is a minimal subsystem of $([a,b]^\ZM,\ZM)$ which is topologically conjugate to the first of the above systems, the rotation action by $\theta$ on $S^1$.
\end{enumerate}
Above we defined three symbolic dynamical systems $\Xi$, $\tilde\Xi$, and $\Xi_\epsilon$, which will play a role in what follows. They are all subsystems of $[a,b]^\ZM$. A strongly pattern equivariant family of Hamiltonians $H_x$ for $x\in [a,b]^\ZM$ defines therefore, by restriction to the three spaces $\Xi$, $\tilde\Xi$, and $\Xi_\epsilon$, families for all these subsystems. We simplify the notation by writing $H_{\phi,t}$ for $H_{\xi(\phi,t)}$ (here $\phi$ is a singular value).

\vspace{0.2cm}

How are these different systems related? We have already commented on various relations between them, but now we want to compare more closely the smoothed out version with the augmented Sturmian subshift. Before doing that let us point out that, from a dynamical point of view and therefore also, as we will see, from a operator algebraic point of view, the smoothed out system $(\Xi_\epsilon,\ZM)$ does not converge to the Sturmian subshift $(\Xi,\ZM)$ when $\epsilon$ tends to $0$. This is already clear from the fact that the space $\Xi_\epsilon$ is homeomorphic to a one dimensional manifold and its sequences have infinite local complexity, whereas $\Xi$ is a Cantor set and its sequences have finite local complexity (the $\xi_n$ take only finitely many values).

%All the above symbolic systems $\Xi$, $\tilde\Xi$, and $\Xi_\epsilon$ are closed invariant subspaces of $[a,b]^\ZM$ and therefore can be compared using the Hausdorff distance $d_H$ on $I( [a,b]^\ZM)$.
%%%%%%%%%%%%%%%%%%%%%%%%%%%
\begin{lemma}\label{lem-eps} With the notation of Section~\ref{sec-prelim}, we have
$$\lim_{\epsilon\to 0} d_H(\Xi_\epsilon,\tilde\Xi) = 0
%\quad \mbox{and}\quad \lim_{\epsilon\to 0} d_H(\hat \Xi_\epsilon,\hat{\tilde\Xi})=0
.$$
\end{lemma}
%%%%%%%%%%%%%%%%%%%%%%%%%
{\bf Proof}: %We prove only the first assertion, as the second can be shown in a similar way. 
Let $ \delta>0$ and $r_\delta$ be the minimal distance between the points of $\{\{n\theta\}|\,|n|\leq \delta^{-1}+1\}\cup\{1\}$. Clearly  $r_\delta\to 0$ if $\delta\to 0$. 
Let $\epsilon_\delta$ be so small that $|\chi(t)-\chi_{\epsilon_\delta}(t)|< \delta$ for all $|t|\geq r_\delta$. We will show that, for all $\epsilon\leq \epsilon_\delta$, within distance $r_\delta+\delta$ of any point in $\Xi_\epsilon$ there is a point in $\tilde \Xi$ and vice versa. 

\vspace{0.2cm}

A point in $\xi(\epsilon,\phi)\in \Xi_\epsilon$ is given by a sequence $\xi_n(\epsilon,\phi) = w_n^\epsilon(\phi)-w_{n-1}^\epsilon(\phi)$
where
$$w_n^\epsilon(\phi)=n \, l_0 - \gamma \big(\varphi_n - \chi_\epsilon(\varphi_n-\theta) \big ) , \ \  \varphi_n = \big\{ \phi + n \theta \big \} .$$
%and $\chi_\epsilon$ is a continuous function which coincides with the indicator function $\chi$ on $\RM^+$ outside the interval $(-\epsilon,\epsilon)$. 
If $\epsilon\leq \epsilon_\delta$ then $|w_n^\epsilon(\phi)- w_n(\phi)|>\delta$ implies that $|\varphi_n-\theta|<r_\delta$ which means that $\phi=\phi'-(n-1)\theta$ for some
$|\phi'|<r_\delta$.
%Let Choose $\epsilon$ so small that $|\chi(t)-\chi_\epsilon(t)|< \delta$ for all $|t|\geq r_\delta$.
Moreover this can happen at most for one $n\in \ZM$ with 
$|n|\leq \delta^{-1}$ and we denote this $n$ by $m$ (if $|w_n^\epsilon(\phi)- w_n(\phi)|\leq \delta$ for all $|n|\leq \delta^{-1}$ then $\xi(\epsilon,\phi)$ is within distance $\delta$ to the sequence $\xi(\phi)$ and we are done). Since $|\phi'|<r_\delta$ we have, for $n\neq m$, $|n|\leq \delta^{-1}+1$, $w_n(\phi) = w_n((1-m)\theta)-\gamma\phi'$.
Define
$$ \tilde w_n = \left\{\begin{array}{ll}
w_n((1-m)\theta) & \mbox{for } n\neq m \\
w_m^\epsilon(\phi)+\gamma\phi' & \mbox{else}
\end{array}\right. $$
Since $(1-m)\theta$ is a singular value and $w_m^\epsilon(\phi)+\gamma\phi'$ lies in between $w_n((1-m)\theta^-)$ and $w_n((1-m)\theta^+)$ the sequence $(\tilde w_n-\tilde w_{n-1})_n$ belongs to $\tilde\Xi$. By construction it is $\delta$-close to the sequence  $(w_n^\epsilon(\phi)-w_{n-1}^\epsilon(\phi))_n$.
%
%(sketchy) Let $\delta>0$ and $\phi\in NS$. Let $(w_n(\phi))_n$ be the sequence of points in $L_\phi\cap (\ZM^2+W)$ as explained above and  $({w^\epsilon}_n)_n$ be the sequence of points in $L_\phi\cap \ZM^2+W_\epsilon$ obtained if we take the smoothed out version for $W$.
%Hence $\xi_n(\phi)=w_n-w_{n-1}$ defines a sequence in $\Xi$ and $\xi_n(\epsilon,\phi)={w^\epsilon}_n-{w^\epsilon}_{n-1}$ one in $\Xi_\epsilon$. 
%Let $D_\epsilon\subset \ZM$ the set of $n$ such that $|w_n-{w^\epsilon}_n| >\delta$. Let $N_\epsilon$ be the minimal separation distance between points in $D_\epsilon$. Since $\phi$ is non-singular, $L_\phi$ does not hit the discontinuity of $W$ and therefore $N_\epsilon \to +\infty$ if $\epsilon\to 0$. We want to find, an $\eta\in\tilde\Xi$ whose distance to $\xi(\epsilon,\phi)$ is smaller than $\delta$. This is the case if, for all $|n|\leq \delta^{-1}$ we have $|\eta_n-{\xi_n(\epsilon,\phi)}| \leq \delta$. If $\epsilon$ is small enough then at most one $n\in D_\epsilon$ satisfies $|n|\leq \delta^{-1}$. Suppose there is one, let's say $m$. Then close to $\phi$ must be a singular value $\phi'$ such that $w_n(\phi) = w_n(\phi')$ for all $|n|\leq \delta^{-1}$ 
%within distance $\delta$ of $x$ must be a singular sequence $x'\in \Xi$ whose flipping point is at $m$. Replacing in $x'$ the $x'_m$ by ${x_\epsilon}_m$ we obtain an element in $\tilde\Xi$ which is $\delta$-close to $x_\epsilon$. This shows that within distance $\delta$ of any point in $\Xi_\epsilon$ there is a point in $\tilde\Xi$.

\vspace{0.2cm}

Consider now $\xi(k\theta,t)\in \tilde\Xi$. It is given by $\xi_n(k\theta,t)= \tilde w_n-\tilde w_{n-1}$ where $\tilde w_n = w_n(k\theta)$ if $n+k\neq 1$ and $\tilde w_{1-k} = (1-k)l_0-\gamma(\theta-t)$.
%If $\epsilon\leq\epsilon_\delta$ then, for all $n,m\neq 1-k$ with $|n|,|m|\leq \delta^{-1}+1$ and all $|\phi'|\leq r_\delta$ we have $|w_n(\theta)-w_{m}(\theta)-(w_n^\epsilon(\theta+\phi')-w_m^\epsilon(\theta+\phi'))|\leq \delta$. 
If $\epsilon\leq\epsilon_\delta$ then, for all $n\neq 1-k$ with $|n|\leq \delta^{-1}+1$ and all $|\phi'|\leq r_\delta$ we have $|w_n(\theta+\phi')-w_n^\epsilon(\theta+\phi'))|\leq \delta$.
On the other hand, if $\phi'$ ranges over $(-r_\delta,r_\delta)$ then $w_{1-k}^\epsilon(\theta+\phi')$ ranges from at least $(1-k)l_0-\gamma(\theta-r_\delta-\delta)$ to $(1-k)l_0-\gamma(\theta+r_\delta-1+\delta)$. This shows that for some $\phi'$ we have
$|\tilde w_{1-k} - w_{1-k}^\epsilon(\theta+\phi')|\leq r_\delta+\delta$. For that $\phi'$
the distance between $\xi(\theta,t)$ and $(w^\epsilon_n(\theta+\phi')-w^\epsilon_{n-1}(\theta+\phi'))_n$ is smaller than $r_\delta+\delta$. 
%Clearly $r_\delta$ tends to $0$ if $\delta$ tends to $0$.
As the set of $\xi(k\theta,t)$ with $k\in\ZM$ and $t\in [0,1]$ lies dense in $\tilde\Xi$, we are done.
%
%where
%Now let $\tilde x\in\tilde \Xi$. Thus there is a singular
%$x\in\Xi$ with flipping point at some $m\in\ZM$ and $(\tilde x)_m$ is obtained from $x_m$ by shifting the point $x_m$ by $t(b-a)$. Let $\phi$ be the intersept (phase) for $x$. If $\epsilon$ is small enough then $|{x_\epsilon(\phi)}_n-x_n|\leq \delta$ for all $|n|\leq \delta^{-1}$ except, perhaps, $n=m$. Moreover, we can move $\phi$ a little bit so as to obtain  ${x_\epsilon(\phi)}_m = \tilde x_m$ without violating this condition. This shows that within distance $\delta$ of any point in $\tilde\Xi$ there is a point in $\Xi_\epsilon$.
\hfill q.e.d.
%%%%%%%%%%%%%%%%%%%%%%%%

\begin{corollary}\label{Cor-BulkEpsilonLimit}
Let $H_x$ be a strongly pattern equivariant Hamiltonian associated to $([a,b]^\ZM,\ZM)$. 
Let $\xi_n\in \Xi_{1/n}$.
Then
$$\lim_{n\to +\infty} \spec(H_{\xi_n}) = \spec(H_{\tilde\Xi}) = \bigcup_{t\in [0,1]} \spec(H_{(\phi,t)})$$
for any singular $\phi$.
%as $\epsilon$ tends to $0$.
\end{corollary}
%%%%%%%%%%%%%%%%%%%%%%%%%%
{\bf Proof}:  $(\Xi_\epsilon,\ZM)$ is minimal and thus all $H_x$, $x\in\Xi_\epsilon$ have the same spectrum and hence $\spec(H_x) = \spec(H_{\Xi_\epsilon})$.
The statement follows therefore from Lemma~\ref{lem-eps} and Theorem~\ref{thm-BBdN16}. 
The last equality follows from the fact that $\spec(H_{(\phi,t)})$ is independent of the choice of the singular value $\phi$, as different choices lead to  unitarily equivalent operators.
\hfill q.e.d.
\bigskip

%%%%%%%%%%%%%%%%%%%%%%%%%%%%%%%%%%%%%%
%%%%%%%%%%%%%%%%%%%%%%%%%%%%%%%%%%%%%%

\section{The winding number of Dirichlet eigenvalues}\label{sec-wind}

It is a generic feature of operators $\hat H_x$ obtained by compression of strongly pattern equivariant operators $H_x$ on $\ell^2(\ZM)$ onto the half space $\ell^2(\NM)$  that their spectrum is richer than that of $H_x$. Indeed, $H_x$ and $\hat H_x \oplus \tilde \Pi^\ast H_x \tilde \Pi$, with $\tilde \Pi$ the partial isometry from $\ell^2(\ZM\setminus \NM)$ to $\ell^2(\ZM)$, differ just by a finite rank perturbation. In this case it is known (see {\it e.g.} \cite{Simon1998}) that ${\rm spec}(\hat H_x \oplus \tilde \Pi^\ast H_x \tilde \Pi) = {\rm spec}(\hat H_x) \cup {\rm spec}(\tilde \Pi^\ast H_x \tilde \Pi)$ contains ${\rm spec}(H_x)$ (which is essential spectrum) plus additional discrete spectrum. As such, if $\Delta$ is a gap in the spectrum of $H_x$, then $\hat H_x$ may contain eigenvalues in that gap, that is, isolated spectral values $\mu$ such that
$$\hat H_x \psi = \mu \psi $$
for $\psi\in\ell^2(\NM)$. The number of eigenvalues in each gap $\Delta$, counted with their degeneracy, cannot exceed the rank of the perturbation \cite{Simon1998}. A square integrable solution $\psi$ to that equation will be referred to as a boundary state, as it is localized near the boundary. Since such a solution can be interpreted as satisfying Dirichlet boundary conditions at $n=0$ we call $\mu$ a Dirichlet eigenvalue. Due to the mirror symmetry, the localized states measured at the mirror symmetric points in the experiments of \cite{BLL} can be theoretically described as boundary states of $\hat H_x$. Of course, $\mu=\mu(x)$ depends on $x$ and we are interested in its behaviour under variation of $x$. 
 %%%%%%%%%%%%%%%%%%%%%%%%%%%%%%%%%%%%%%%%%%%
 \begin{corollary}\label{cor-cont}
Let $X$ be any of the three $\Xi$, $\Xi_\epsilon$, or $\tilde \Xi$. 
The map $X \ni x \mapsto \spec(\hat H_x)$ is continuous. 
\end{corollary} 
%%%%%%%%%%%%%%%%%%%%%%%%%%%%%%%%%%%%%%%%%%
{\bf Proof}: In the cases in which $X$ is $\Xi$ or $\Xi_\epsilon$ this is a restatement of Corollary~\ref{cor-cont1} as both systems are minimal. 
In the third case  $\tilde \Xi$ is not minimal, but $\Xi$ is its minimal component. As any $\hat H_{(\phi,t)}$ is a finite rank perturbation of $\hat H_{(\phi,0)}$ and $\hat H_{(\phi,1)}$ continuity in $t$ follows from analytic perturbation theory. The continuity extends therefore from $\Xi$ to  $\tilde \Xi$. \hfill q.e.d.
\medskip

For $\Xi$, which is a Cantor set, the continuity proven above does not exclude singular behavior of the boundary spectrum. 
%Indeed, let us restrict to the model \eqref{eq-model}, in which case the perturbation mentioned above is of rank 2. 
Indeed, let $\Delta$ be a spectral gap of $H_\Xi$, {\it i.e.} a connected component of the resolvent set of $H_\Xi$. Consider a closed interval $\Delta' \subset \Delta$.
Pick $\xi$ arbitrary from $\Xi$. Then ${\rm spec}(\hat H_\xi) \cap \Delta'$ consists of at most finitely many isolated points; let $K$ be an upper bound to their number. Let $\mu(\xi)$ be such a point, if it exists at all, and let $\psi$ be its normalized eigenvector. Since the distance of $\Delta'$ to the spectrum of $H_\Xi$ 
is strictly positive,
a direct application of Thomas-Combes theory shows that $\psi(n) \leq c e^{-\beta n}$ with real constants $c$ and $\beta>0$ which do not depend on 
%$ \delta=d\big (\Delta',{\rm spec}(H_\Xi)\big )$ 
$\xi$ (as long as $\mu(\xi)\in\Delta'$).

\vspace{0.2cm}

Consider now 
%$N$ forward windings of $L_\theta$ from Fig.~\ref{Fig-Sturmian}. 
the first $N$ singular points $\phi_n = n\theta$, $i=1,\cdots N$. 
They divide the circle $S^1$ into $N$ connected open sets and consequently $\Xi$ into $N$ clopen sets $I_j $.  Any two sequences from the same $I_j$ agree on their first $N$ entries.
The $\xi$ chosen above necessarily falls into one of these, say $I_k$.
Thus, for any $\xi' \in I_k$ such that $\mu(\xi')\in\Delta'$ we have $\big ( \hat H_{\xi'}-\mu(\xi) \big ) \psi (n) = 0$ provided $n<N-r$ where $r$ is the sum of the range of $H_\xi$ and the largest range of the sliding block codes. It follows that  
$$
\big \| \big ( \hat H_{\xi'}-\mu(\xi) \big ) \psi\|\leq 2\|H_\xi\| \left(\sum_{n\geq N-r} \|\psi(n)\|^2\right)^\frac12 \leq c' e^{-\beta N}
$$
for some constant $c'$ which again, does not depend on $\xi$.
This implies that the distance between $\mu(\xi)$ and the spectrum of $\hat H_{\xi'}$ is bounded from above by $c' e^{-\beta N}$. We conclude that ${\rm spec}(\hat H_{I_k}) \cap \Delta'$  is contained in (at most) $K$ intervals of widths less than $c'e^{-\beta N}$. As such, ${\rm spec}(\hat H_\Xi) \cap \Delta' = \bigcup_{k=1}^N{\rm spec}(\hat H_{I_k}) \cap \Delta'$ is contained in at most $KN$ 
%possibly overlapping 
intervals of widths at most $c'e^{-\beta N}$. Hence, the Lebesque measure of ${\rm spec}(\hat H_\Xi) \cap \Delta'$ cannot exceed $c' KN e^{-\beta N}$. But this upper bound goes rapidly to zero as $N \rightarrow \infty$. We have just proved:

\begin{proposition}\label{pro-KB} The spectrum ${\rm spec}(\hat H_\Xi) \cap \Delta$ has Lebesque measure $0$. In particular it contains gaps.
\end{proposition}

Because of the last result, we cannot define a spectral flow through the gap as $x$ varies in $\Xi$, and neither a winding number. Consider, however, the situation in which $X$ is homeomorphic to the circle $S^1$ and there is one Dirichlet eigenvalue $\mu(x)$ in the gap for some $x\in X$.
The continuity of $x \mapsto \spec(\hat H_x)$ implies that Dirichlet eigenvalues depend continuously on $x$. Therefore  the eigenvalue $\mu(x)$ cannot disappear in the interior of the gap under variation of $x$ but only be absorbed by the gap-boundaries or have a trajectory that closes into itself without touching the gap-boundaries. Identifying the gap boundaries, the gap becomes a circle and the function $x\mapsto \mu(x)$ defines a winding number (with the convention that, if there is no eigenvalue at $x$ inside the gap $\Delta$, then $\mu(x)$ coincides with the gap boundary). Now if there are several Dirichlet eigenvalues they define several continuous functions $x\mapsto \mu_i(x)$ and the winding number of the gap $\Delta$ is the sum of the winding numbers defined by the individual curves in $\Delta$. Strictly speaking, if there is level crossing then we have to make choices in the definition of these functions, but the sum of their winding numbers can be easily seen to be independent of these choices. Stated differently, the winding number of gap $\Delta$ depends only on the spectral flow of the boundary states in $\Delta$, that is, the restriction of $x \mapsto \spec(\hat H_x)$ to the gap.

%%%%%%%%%%%%%%%%%%%%%%%%%
\section{Numerical Experiments}\label{sec-numerics}
%%%%%%%%%%%%%%%%%%%%% 
In this section we present the results of the numerical calculations of various spectra. We will restrict to one specific Hamiltonian of the type \eqref{eq-model}:
\begin{equation}\label{Eq-TheHamiltonian}
H_\xi = T + T^\ast + 2 \tilde b_\xi, \quad \xi \in [a,b]^\ZM,
\end{equation}
where the sliding block code is $b:[a,b] \rightarrow \RM$, $b(t)=(t-a)/(b-a)$. The above Hamiltonian will also be subjected to Dirichlet boundary conditions. We follow the notation from Section~\ref{sec-prelim}, hence $H_\Xi$ denotes the family $\{H_\xi\}_{\xi \in \Xi}$ etc.\,. Standard numerical approximants are employed and the targeted accuracy was such that the numerical errors will not be detectable by the eye in the plots reported here. The value of $\theta$ has been fixed to $\frac{3-\sqrt{5}}{2}$, hence we simulate Fibonacci sequences.

\subsection{Bulk spectra} 
\begin{figure}[H]
\center
\includegraphics[width=\textwidth]{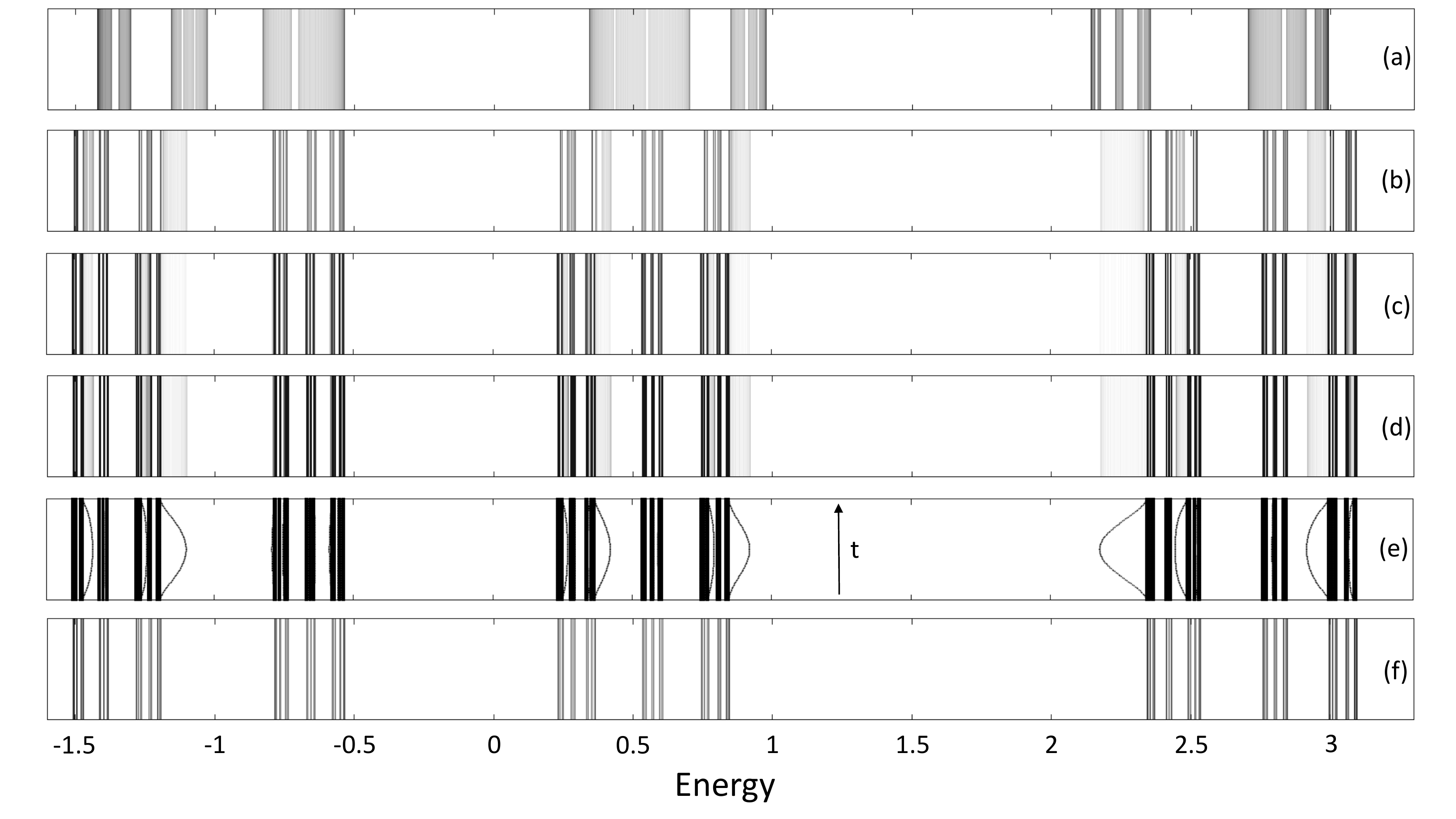}
\caption{{\small Bulk spectra of $H_X$ for the three choices for $X$. The first three panels show $\spec(H_{\Xi_\epsilon})$ with $\epsilon=0.1$ (a), $\epsilon=0.01$ (b), $\epsilon=0.001$ (c). Panel (d) shows $\spec(H_{\tilde\Xi})$ and panel (e) how $\spec(H_{\phi,t})$ depends on $t\in [0,1]$. Finally, panel (f) shows   $\spec(H_{\Xi})$. The computations were performed with periodic boundary conditions on a finite system of size $6765$, which came from the rational approximation: $\theta=\frac{3-\sqrt{5}}{2} = \frac{2584}{6765}-9.77 \times 10^{-9}$.}}
\label{Fig-BulkSpectra}
\end{figure}

In Fig.~\ref{Fig-BulkSpectra} we show the bulk spectra for the models associated to different spaces. The bulk spectra are computed with periodic boundary conditions on a finite approximant.
The figure confirms and exemplifies the statement of Cor.~\ref{Cor-BulkEpsilonLimit}. Indeed, it shows clearly that ${\rm spec}(H_{\Xi_\epsilon})$ converges to ${\rm spec}(H_{\tilde \Xi})$ in the limit $\epsilon \rightarrow 0$, and not to ${\rm spec}(H_\Xi)$. 
We recall that, whereas ${\rm spec}(H_{\Xi_\epsilon}) = {\rm spec}(H_{\xi_\epsilon})$ for any point $\xi_\epsilon \in \Xi_\epsilon$ and ${\rm spec}(H_{\Xi}) = {\rm spec}(H_{\xi})$ for any point $\xi \in \Xi$, we have ${\rm spec}(H_{\tilde\Xi}) = \bigcup_{t\in [0,1]} {\spec}(H_{\phi,t})$ for any singular choice of $\phi$.
While the numerical data show that the prominent spectral gaps in $\spec(H_\Xi)$ remain open when augmenting to $\tilde \Xi$, this is not the case for all of the spectral gaps. We will show however below, that the exponential map of the six-term exact sequence associated to the left SES of diagram \ref{eq-dia} is trivial, which is a sign that the closing of these gaps is not topological (the $K$-group elements defined by the spectral gaps in ${\rm spec}(H_\Xi)$ lift to $K$-group elements in the $K_0$-group for the augmented model).

%The statements obtained from $K$-theory are on the spectra of $\hat H_{X}$, $X=\Xi,\tilde \Xi, \Xi_\epsilon$. 

\subsection{Boundary spectra and Dirichlet eigenvalues}

The set 
$$
\spec_b(X)=\bigcup_{x\in X}\spec(\hat H_x) \setminus \spec(H_X) %\quad X=\Xi,\tilde \Xi, \Xi_\epsilon,
$$
may be referred to as the boundary spectrum, as all its values correspond to eigenstates localized at the boundary. We call the boundary spectrum topological if it covers a bulk gap completely, regardless of the imposed boundary conditions. This property, while  interesting in itself, is a necessary condition for a topologically non-trivial spectral flow at the boundary. 

\vspace{0.2cm}

To visualize the spectra of $\hat H_\xi$, $\xi\in X$ where $X$ is $\Xi$ or $\Xi_\epsilon$, we 
first note that the (uncountable) union over all space $\bigcup_{\xi\in X} \spec(\hat H_\xi)$ yields the same as the countable union along a single orbit
$\bigcup_{n\in\ZM} \spec(\hat H_{\alpha^n(\xi)})$ where $\xi$ can be any choice. This follows from the fact that the systems $(X,\ZM)$, $X= \Xi$ or $\Xi_\epsilon$, are forward minimal (every forward orbit is dense). Now we approximate 
$\spec(\hat H_\xi)$ by  
%$\bigcup_{n\in\ZM} \spec(\hat H_{\alpha^n(\xi)})$ by $\bigcup_{n=0}^{L-1}
$\spec \Big (H^{(D)}_{\xi_L} \Big )$ where $\xi_L =\{\xi_0,\ldots \xi_{L-1}\}$ is the restriction to the finite length $L$ part of $\xi$ and we impose Dirichlet conditions at the two boundaries. 
Of course, $\spec \Big (H^{(D)}_{\xi_L} \Big )$ includes the Dirichlet eigenvalues of both boundaries, that is the eigenstates localized at the left and at the right boundary. But due to the inversion symmetry of Sturmian sequences, namely that $\xi_n\mapsto \xi_{-n}$ preserves the spaces $\Xi$ and 
$\Xi_\epsilon$, the spectrum coming from the left Dirichlet eigenvalues and that coming from the right Dirichlet eigenvalues have to coincide and thus 
%spectrum 
%we expect $\bigcup_{\xi\in X} \spec(H_{\xi_L}^{(D)})$ to well approximate $\bigcup_{\xi\in X} \spec(\hat H_\xi)$. 
%Furthermore, since the systems $(X,\ZM)$, $X= \Xi$ or $\tilde\Xi$ are forward minimal (every forward orbit is dense) the (uncountable) union over the space $\bigcup_{\xi\in X} \spec(\hat H_\xi)$ yields the same as the countable union along a orbit
%$\bigcup_{n\in\ZM} \spec(\hat H_{\alpha^n(\xi)})$, for any choice of $\xi\in\X$.  
%$\bigcup_{n=0}^{L-1}\spec \Big (H^{(D)}_{\alpha^n(\xi)_L} \Big )$ 
$\bigcup_{n=0}^{L-1}\spec \Big (H^{(D)}_{\alpha^n(\xi_L)} \Big )$
approximates $\bigcup_{\xi\in X} \spec(\hat H_\xi)$ where the index $k$ in the
shift action $\alpha(\xi_k) = \xi_{k+1}$ is to be taken modulo $L$. 
\begin{figure}[H]
\center
\includegraphics[width=\textwidth]{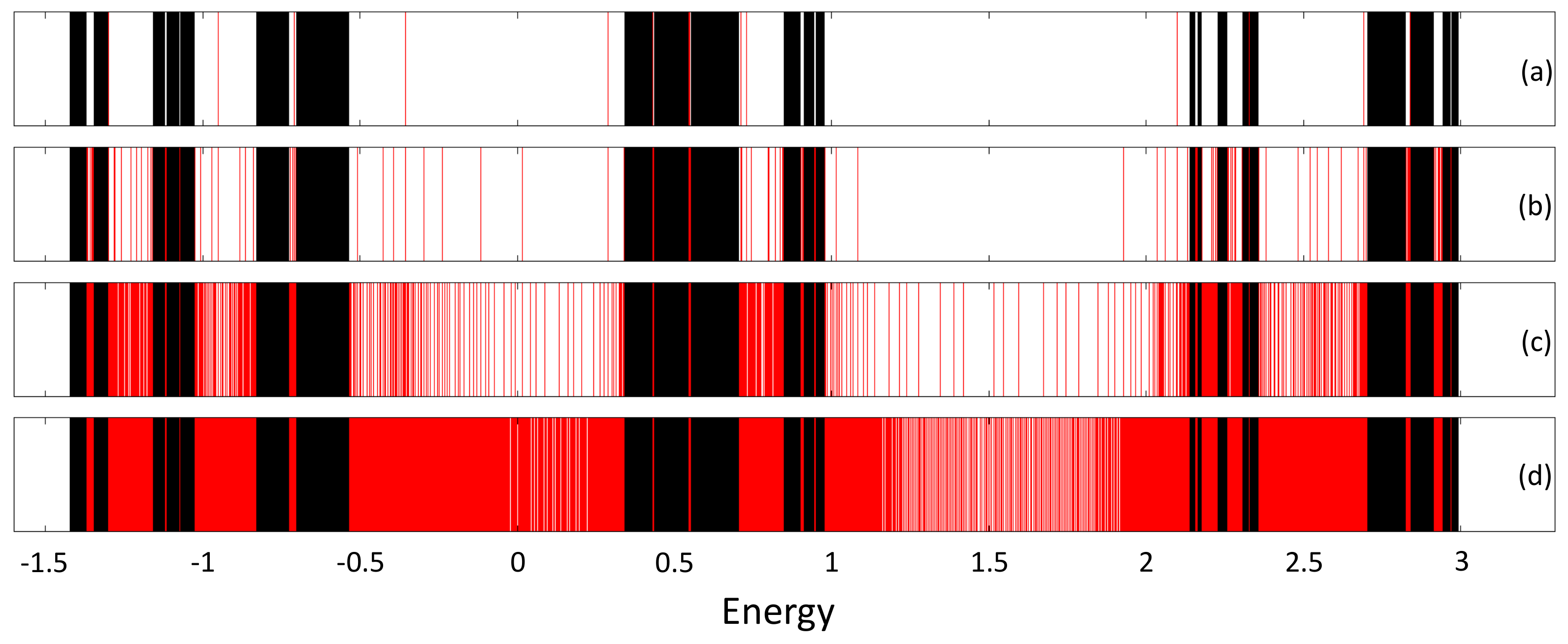}
\caption{{\small Boundary spectrum $\spec_b (\hat H_{\Xi_\epsilon})$, $\epsilon=0.1$ (shown in red). For reference, the spectrum of the periodic model from Fig.~\ref{Fig-BulkSpectra}(b) has been overlapped and shown in black. The panels show $\bigcup_{n}\spec \Big (H^{(D)}_{\alpha^n(\xi_L)} \Big )$ for one $\xi \in \Xi_\epsilon$ ($L=6765$ as in Fig.~\ref{Fig-BulkSpectra}), with (a) $n=0$, (b) $n=0,\ldots,9$, (c) $n=0,\ldots,99$ and (d) $n=0,\ldots,999$.}}
\label{Fig-EpsEdgeSpectraEps0p1}
\end{figure}

Fig.~\ref{Fig-EpsEdgeSpectraEps0p1} shows that the boundary spectrum $\spec_b(H_{\Xi_\epsilon})$ ($\epsilon=0.1$) is topological. Indeed, as higher $n$ is considered, i.e.\ more spectra are overlapped, the union of the Dirichlet spectra is seen to sample better and better the bulk gaps. With the present resolution, some of the bulk gaps already appear completely filled in Fig.~\ref{Fig-EpsEdgeSpectraEps0p1}(d). 
In practice, 
%The practical value of this phenomenon resides in the fact that 
the union of the Dirichlet spectra can be obtained by bundling together many (finite) chains
and measuring the states which are localized at one edge. The energies of these states 
fall insight the bulk gap thus creating a strictly positive density of states of the bundle at its boundary. 
%{\tt practically all chains are finite, so is this bundling up with two different kind of boundry conditions, or does one do the mirroring trick?}. 
%In this way one generates states inside the bulk gaps which are localized at the open end of the bundle and have strictly positive density of states. 
These states cannot be moved out of the gap by wearing and tearing, because their existence follows from topological ($K$-theoretic) arguments which are
not conditioned by any particular boundary condition. 
Furthermore, the bundles can be generated simply by cutting a single pattern at consecutive places, as done in Fig.~\ref{Fig-EpsEdgeSpectraEps0p1}. We have verified numerically the these conclusions remain unchanged when $\epsilon$ is reduced even by orders of magnitude.

\vspace{0.2cm}

\begin{figure}[H]
\center
\includegraphics[width=\textwidth]{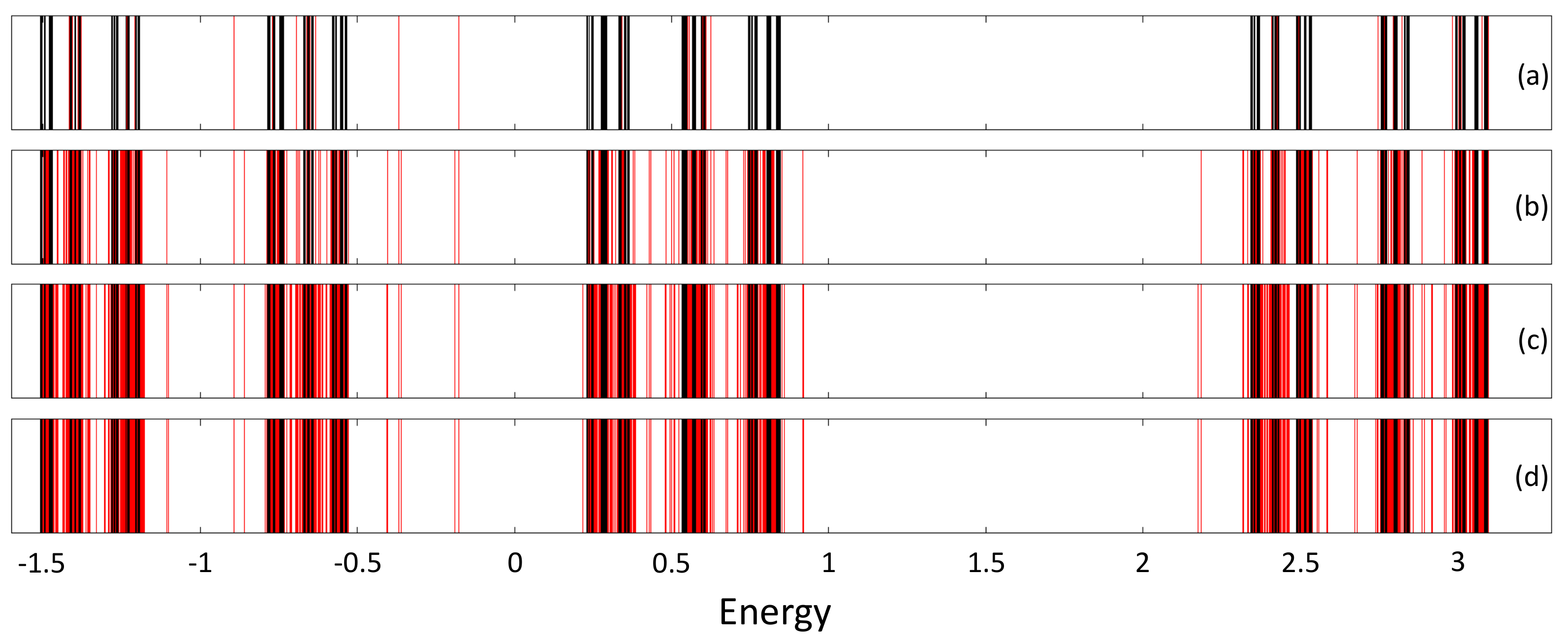}
\caption{{\small The boundary spectrum of the Kohmoto model with Dirichlet boundary condition (shown in red). For reference, the spectrum of the periodic model from Fig.~\ref{Fig-BulkSpectra}(f) has been overlapped and shown in black. The panels show $\bigcup_{n}\spec (H^{(D)}_{\alpha^n( \xi_L)})$ with, as in Fig.~\ref{Fig-BulkSpectra}, $L=6765$  and, for (a) $n=0$, (b) $n=0,\ldots,9$, (c) $n=0,\ldots,99$ and (d) $n=0,\ldots,999$.}}
\label{Fig-EdgeSpectra}
\end{figure}

Fig.~\ref{Fig-EdgeSpectra} shows that the situation is very different for the Kohmoto model.
%reports $\spec_b(H_\Xi)$ of the Kohmoto model and there is stark contrast when the data is compared with Fig.~\ref{Fig-EpsEdgeSpectraEps0p1}. 
We have verified that, even when we take the maximal value for $n$ that is allowed by the finite size of the system, the bulk gaps remain mostly empty of boundary spectrum, in agreement with Proposition~\ref{pro-KB}. This already indicates that the Kohmoto model does not support topological boundary spectrum. %Contrary, the augmented model does support it, as we shall see in the next section.

\subsection{Dependence of the Dirichlet eigenvalues on the intercept}\label{sec-specflow}

\begin{figure}
\center
\includegraphics[width=0.9\textwidth]{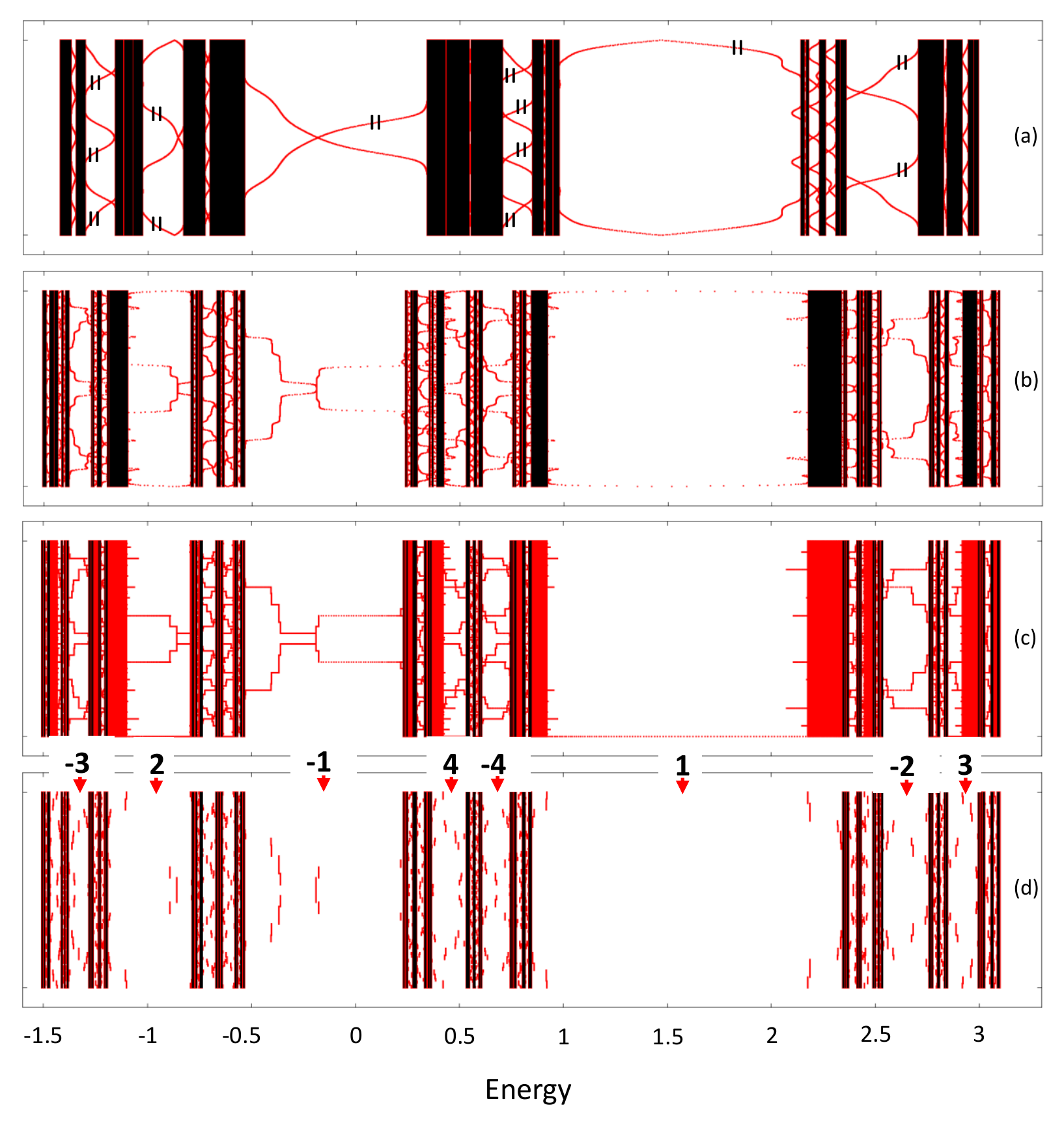}
\caption{{\small Edge spectra rendered as functions of the intercept $\phi$. Panels (a) and (d) re-plot the data from Figs.~\ref{Fig-EpsEdgeSpectraEps0p1}(d) and \ref{Fig-EdgeSpectra}(d), respectively, (b) is same as panel (a) but for $\epsilon=0.01$, while (c) is a plot of $\cup_{t \in [0,1]}\spec(\hat H_{\phi,t})$. The labels of the most prominent gaps (see next figure) are included for the reader's convenience. The marks in panel (a) indicate which bands are localized at the left edge.}}
\label{Fig-EdgeSpectralFlow}
\end{figure} 

Whereas in Figs.~\ref{Fig-EpsEdgeSpectraEps0p1} and \ref{Fig-EdgeSpectra} we have superimposed the spectra of $H_\xi^{(D)}$ for different $\xi \in X$, in Fig.~\ref{Fig-EdgeSpectralFlow} we show their dependence on the intercept $\phi$. Note that $\phi$ is along the $y$-axis and periodic. 
We refer to the red lines in the panels as spectral flow lines, along the parameter $\phi$; they may also be understood as band functions for the boundary spectrum with $\phi$ playing the role of quasi-momentum. It should be kept in mind, that the  curves show the left and right Dirichlet eigenvalues, together. 
However, by modifying the potential at one end of the chain and observing which lines have been affected, we can easily decide which values correspond eigenstates localized at the left and which at the right boundary. Indeed, our observation is that half of the lines remain virtually un-affected, and so we can be sure that the boundary spectra at the two edges are entirely decoupled. This being said, we can focus on the spectral flow of the states localized at the left boundary, which have been marked in panel (a) of the figure (and hence can be also identified in the rest of the panels). 

\vspace{0.2cm}

We now analyze Fig.~\ref{Fig-EdgeSpectralFlow} more closely. As alluded to  already in the previous section, the smooth models shown in panels (a) and (b) display spectral flow lines visible through the red lines 
%chiral spectral flows, visible here through the spectral bands 
connecting the lower and upper gap-boundaries. 
%These are quite different from ``quadratic'' bands which originate and terminate on the same gap-boundary, or from ``flat bands'' which close into themselves without touching the gap-boundaries. 
If we identify the gap-boundaries as in Section~\ref{sec-wind}, then the flow lines wind once or several times around the gap, which is now a circle, and we can easily read-off their winding numbers. 
The data confirm that the spectral flow of the boundary states of the smooth models converge in the $\epsilon \rightarrow 0$ limit to that of the augmented model, shown in panel (c). For this case too, we can define without any difficulty a winding number. But note that in panel (c) the lines are piecewise either horizontal or vertical. The horizontal lines are, of course, not functions of $\phi$ but they arrise as we take the union over $t$. The non-horizontal pieces of the red lines in panel (c) look vertical to the eye and thus the winding number depends only on the horizontal ($t$-dependent) part of the lines.
We exploit this to give an interpretation of the winding number in Section~\ref{sec-8.1}.

\vspace{0.2cm}

We can clearly see in panel (d) that the spectral flow lines are discontinuous as a function of $\phi$ for the Kohmoto model and therefore there is no proper way of defining winding numbers. This picture ought to be compared with Figure~3 (a),(c) of \cite{BLL}, except that in \cite{BLL} $\phi$ is plotted along the $x$-axis. The vertical pieces of the curves of \cite{BLL} Fig.~3 (c) remain unexplained and there is no way to guess them, as is clear, for instance, if one considers the gap with label $1$ of our panel (d). How should one conclude from panel (d) that the winding number of the edge states in that gap should be $-1$?

\vspace{0.2cm}

For the panels (a), (b) and (c), where the winding numbers are well defined, the winding numbers corresponding to the prominent gaps can be seen to coincide with the so called gap labels provided by the $K$-theory and listed above panel (d).

%We see clearly that, when $X=\Xi$ there is no proper way of lining up the vertical red line pieces to continuous curves joining the bulk spectrum. On the other hand, if   $X=\Xi_\epsilon$ this is clearly the case. Furthermore, the figure suggests clearly that if $\epsilon$ goes to $0$ the (red) curves of the Dirichlet eigenvalues approach the curves $\phi \mapsto \bigcup_{t\in[0,1]} \spec (\hat H_{(\phi,t)})$. 

%It should be kept in mind, that the (red) curves in the panels show the left and right Dirichlet eigenvalues together as a function of $\phi$. Furthermore, in the figure $\phi$ is along the $y$-axis and periodic. Without separating left and right dirichlet eigenvalues it is difficult to deduce the winding number we are after (but the curves are compatible with our theorem). 
%{\tt Can we separate left from right Dirichlet eigenvalues? I would put the (d) panel first.}

\begin{figure}
\center
\includegraphics[width=0.9\textwidth]{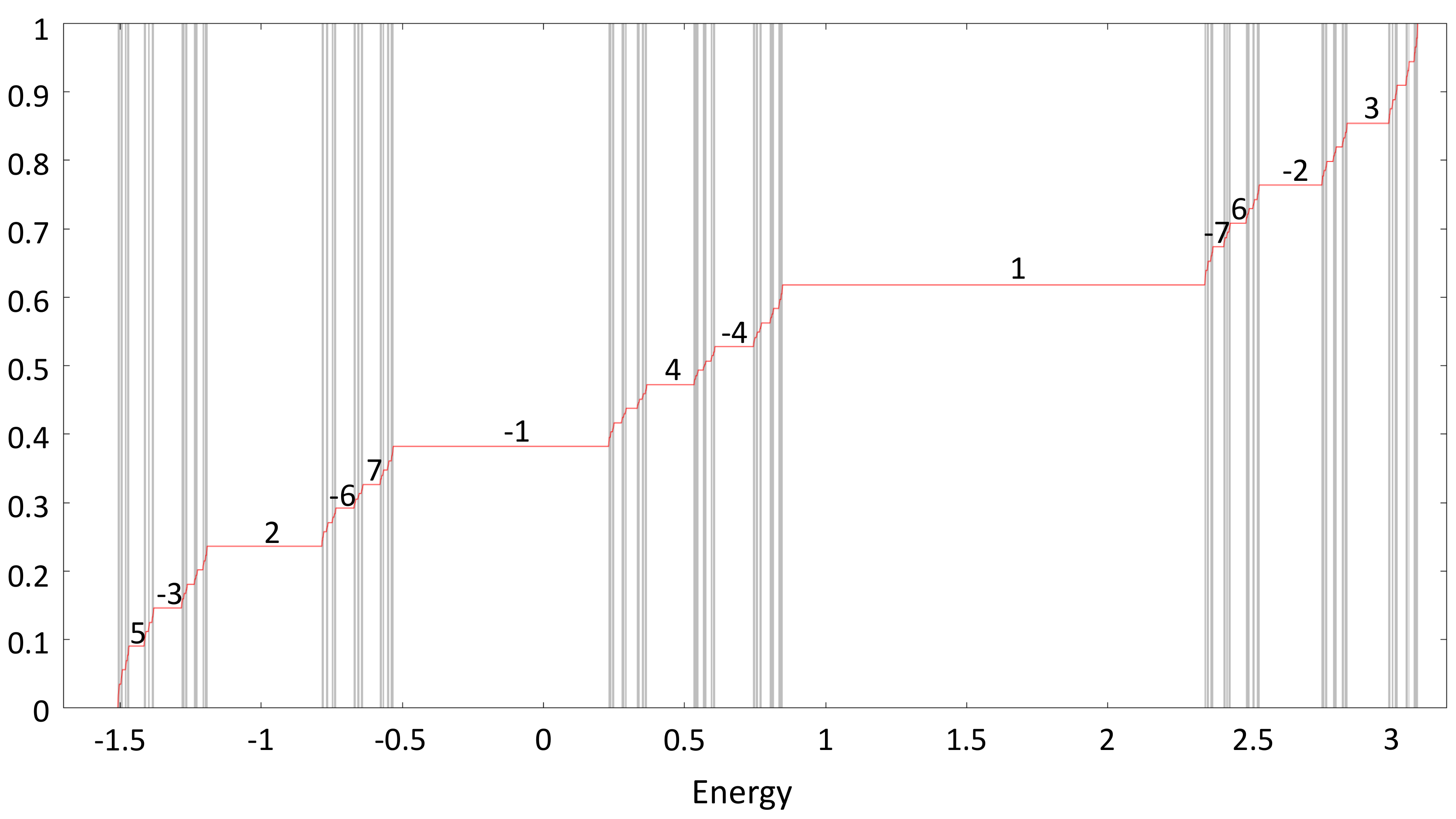}
\caption{{\small The spectrum $\spec(H_\Xi)$ (shaded regions), replotted from Fig.~\ref{Fig-BulkSpectra}, and its integrated density of states (red curve). Several prominent spectral gaps can be observed, together with their labels.}}
\label{Fig-GapLabeling}
\end{figure}

\subsection{Integrated density of states at gap energies}

Fig.~\ref{Fig-GapLabeling} shows a numerical representation of the integrated density of states (IDS) for the Kohmoto model. It has been computed as
\begin{equation}\label{Eq-IDS}
IDS(E) = \frac{\mbox{\rm number of eigenvalues of $H_{\xi_L}$ smaller than $E$}}{L},
\end{equation}
where $L$ is the finite size of the periodic approximant used in the simulations. For Kohmoto model, it is known that the IDS, when evaluated inside the spectral gaps, can take only the values \cite{Bel86,Bel95}
\begin{equation}\label{Eq-IDSValues}
IDS \subseteq \{n+m\theta, \ n,m \in \ZM\} \cap [0,1].
\end{equation}
As we shall see, the labels $(n,m)$ are bulk topological invariants supplied by the $K$-theory of the algebra of bulk observables. Fig.~\ref{Fig-GapLabeling} also reports the label $m$ of the prominent gaps, as computed from the numerical IDS. One central issue in our work is how to relate this bulk topological invariant to the topological edge spectrum. Note that all gap labels $m$ in Fig.~\ref{Fig-GapLabeling} are different from zero, which is a direct consequence of \eqref{Eq-IDSValues}. Indeed, the only gap labels $(n,m)$ with $m=0$ for which $IDS$ fits inside the interval $[0,1]$ are $n=0$ or $1$, but these cases correspond to the empty or fully populated spectrum, respectively.

\section{$C^*$-algebras}
\label{sec-calgebras}
Our proof of the bulk-boundary correspondence uses the $C^*$-algebraic description of covariant families of Schr\"odinger operators. 
\subsection{Preliminaries}
Two $C^*$-algebras associated to 
a topological dynamical system given by a homeomorphism 
$\alpha$ on a compact space $X$ will be of importance. 
The first one is the crossed product algebra $C(X)\rtimes_\alpha\ZM$. It is the universal $C^*$-closure of the $*$-algebra $C(X)_\alpha\ZM$ given by finite sums of finite products of elements of $C(X)$ and a unitary $u$ such that $u f u^* = f\circ\alpha$. Each 
$x\in X$ induces a representation $\pi_x$ of $C(X)_\alpha\ZM$ on $\ell^2(\ZM)$,
$$\pi_x(f) \psi (n) = f(\alpha^n(x))\psi(n),\quad \pi_x(u)= T $$
where $f\in C(X)$ and $T$ is the left translation operator, $T\psi(n) = \psi(n+1)$. 
The universal $C^*$ norm on $C(X)_\alpha\ZM$ is given by $\|H\|:=\sup_{x\in X} \|\pi_x(H)\|$. 

In the context of symbolic dynamics where $X$ is a symbolic subshift, a sliding block code $b$, extended to $C(X)$, defines an element of $C(X)_\alpha\ZM$ and 
$\pi_x(b)=\tilde b_x$ as defined in (\ref{eq-b1}). Thus the covariant family (\ref{eq-b2}) coincides with the family $(\pi_x(H)_{x\in X}$ where 
$H = \sum_{k\in S} (b_k u^k + u^{-k} \bar{b_k})$. The covariant family is thus described by a single element in the $C^*$-algebra and, since the family of representations $\pi_x$, $x\in X$ is faithful, this yields a one-to-one correspondance. A single representation $\pi_x$ is faithful if the orbit of $x$ is dense, and then we may identify  $C(X)\rtimes_\alpha\ZM$ also with the norm closed subalgebra of $\Bb(\ell^2(\ZM))$ of that representation.

A second algebra associated to the dynamical system is the Toeplitz extension algebra $\Tt(C(X),\alpha)$. It is the universal $C^*$-closure of the $*$-algebra $C(X)_\alpha\NM$ given by finite sums of finite products of elements of $C(X)$ and a proper coisometry $v$ such that $v f v^* = f\circ\alpha$. $v$ being an coisometry means that $vv^*=1$ and $v^*v=1-\hat e$ where $\hat e$ is a non-zero projection. 
%The relations imply that $v^* f v = (1-e)f\circ\alpha^{-1}$. 
Each 
$x\in X$ induces a representation $\hat\pi_x$ of $C(X)_\alpha\NM$ on $\ell^2(\NM)$,
$$\hat\pi_x(f) \psi (n) = f(\alpha^n(x))\psi(n),\quad \hat\pi_x(v)= \hat T $$
where $\hat T$ is the left translation operator on $\ell^2(\NM)$, i.e.\ $\hat T\psi(n) = \psi(n+1)$. Again, the family of representations $\hat\pi_x$, $x\in X$ is faithful.
Note that $\hat T$ is the compression of $T$ to $\ell^2(\NM)$ and hence, in the context of symbolic subshifts, our family of half-space operators (\ref{eq-hs2}) is faithfully represented by the element $\hat H = \sum_{k\in S} (b_k v^k + {v^*}^k \bar{b}_k)\in \Tt(C(X),\alpha)$. 
%%
%%Again, the family of representations $\hat\pi_x$, $x\in X$ is faithful and thus an element $\hat H\in \Tt(C(X),\alpha)$ faithfully represented by $\hat\pi_X(\hat H)$ (and
%%the universal $C^*$ norm on $C(X)_\alpha\NM$ is given by $\|\hat H\| = \sup_{x\in X} \|\hat\pi_x(\hat H)\|$. 
%Now a single representation $\hat\pi_x$ is faithful if the forward orbit $\{\alpha^n(x)\,|\, n\in\NM\}$ of $x$ is dense and then we may identify  $\Tt(C(X),\alpha)$ with the norm closed subalgebra of $\Bb(\ell^2(\ZM))$ of that representation. 
%%$\hat\pi_x(\hat H) = \Pi\pi_x(H)\Pi$ where $\Pi$ is the orthogonal projection onto $\ell^2(\NM)$. 
%
%\noindent *******Emil******
%
%A single $\hat \pi_x$ representation cannot be faithful. Indeed:
%\begin{equation}
%\hat \pi_x(f \hat e) = f(x) |0\rangle \langle 0 |.
%\end{equation}
%But there are plenty of continuous functions $f$ over $X$ which just happen to vanish at a particular $x \in X$. In such cases, $f\, \hat e \neq 0$ in $\Tt(C(X),\alpha)$ but $\hat \pi_x(f\, \hat e) =0$.
%********Emil************
\newcommand{\Tp}{\pi}
\newcommand{\Ti}{i}

The map $\hat e\mapsto 0$ induces a unital surjective $*$-algebra morphism 
$C(X)_\alpha\NM\to C(X)_\alpha\ZM$ which extends to a surjective $*$-algebra morphism $\Tp:\Tt(C(X),\alpha)\to C(X)\rtimes_\alpha\ZM$ whose kernel can be seen to be isomorphic to $C(X)\otimes \Kk$. The associated short exact sequence 
\begin{equation}\label{eq-TE}
0 \to  C(X)\otimes\Kk  \stackrel{\Ti}\to  \Tt(C(X),\alpha)  \stackrel{\pi}\to  C(X)\rtimes_\alpha \ZM  \to 0
\end{equation}
is called the Toeplitz extension. 

\subsection{An augmented extension for Sturmian systems}
In the context of Sturmian systems we are interested in the case 
%the crossed product $C(X)\rtimes_\alpha\ZM$ and its Toeplitz extension $\Tt(C(X),\alpha)$ play a role for the symbolic dynamical systems 
%which we obtain 
where $X$ is $\Xi$ or $\tilde \Xi$, and the action is induced by the left shift.
The inclusion of $\Xi$ into $\tilde\Xi$ induces a surjection $C(\tilde\Xi)\to C(\Xi)$ which commutes with the $\ZM$-action and thus gives rise to a surjection
$q:C(\tilde\Xi)\rtimes_\alpha\ZM \to C(\Xi)\rtimes_\alpha\ZM$.
For the bulk-boundary correspondance which we discuss below we compose the two surjective maps, $\Tt(C(\tilde\Xi),\alpha)\to C(\tilde\Xi)\rtimes_\alpha\ZM \to C(\Xi)\rtimes_\alpha\ZM$ to obtain the {\it augmented} exact sequence
\begin{equation}\label{eq-JE}
0 \to  J \hookrightarrow  \Tt(C(\tilde\Xi),\alpha)  \stackrel{\tilde \Tp}\to  C(\Xi)\rtimes_\alpha \ZM  \to 0
\end{equation}
where $J$ is by definition the kernel of $\tilde\Tp := q\circ\Tp$. To better understand $J$ consider the following diagram of short exact sequences in which all squares commute
\begin{equation}\label{eq-dia}
\begin{array}{rcccccl}
&  &  &  &  &  0 & \\
&  & & & &\downarrow  & \\
& 0 &  &  & & C_0(\Ss)\rtimes_\alpha\ZM & \\
& \downarrow & & & & \downarrow & \\
0 \to & C(\tilde \Xi)\otimes\Kk & \stackrel{\Ti}\to & \Tt(\tilde\Xi,\alpha) & \stackrel{\pi}\to & C(\tilde\Xi)\rtimes_\alpha \ZM & \to 0\\
& \downarrow \Ti & & || & & \downarrow q & \\
0 \to & J & \hookrightarrow & \Tt(\tilde\Xi,\alpha) & \stackrel{q\circ\pi}\to & C(\Xi)\rtimes_\alpha \ZM & \to 0\\
& \downarrow \pi& & & &\downarrow  & \\
& C_0(\Ss)\rtimes_\alpha\ZM  &  &  &  &  0 & \\
& \downarrow & & & &  & \\
& 0 &  &  &  &  &
\end{array}
\end{equation}
The first horizontal SES is the Toeplitz extension (\ref{eq-TE}) for $X=\tilde\Xi$. The right vertical SES comes from the inclusion $\Xi\subset\tilde\Xi$ which induces the surjection $q:C(\tilde\Xi)\rtimes_\alpha\ZM\to C(\Xi)\rtimes_\alpha\ZM$. With $\Ss:=\tilde\Xi\backslash\Xi$ the ideal in that sequence is just $C_0(\Ss)\rtimes_\alpha\ZM$. 

The second horizontal SES defines $J$ as the kernel $J=\ker q\circ\pi$. A diagram chase shows
that the left vertical sequence is short exact: Let $x\in \Tt(\tilde\Xi,\alpha)$ then $x\in J$ iff $q(\pi(x))=0$ iff $\pi(x)\in C_0(\Ss)\rtimes_\alpha\ZM$.
Let $E$ be the preimage under $\pi$ of $C_0(\Ss)\rtimes_\alpha\ZM$, then  
$\pi(x)\in C_0(\Ss)\rtimes_\alpha\ZM$
iff $\exists y\in E:\pi(y) = \pi(x)$ iff $x \in E + I$ where $I = i(C(\tilde \Xi)\otimes\Kk)$. Hence $J = E+I$. It follows that $J/I = E/(E\cap I)=E/\ker\pi \cong C_0(\Ss)\rtimes_\alpha\ZM$. 
%
%{\it Our aim  is to compute the boundary correspondance which arises from the second horizontal SES 
%\begin{equation}\label{eq-JE}
%0 \to  J \hookrightarrow  \Tt(\tilde\Xi,\alpha)  \stackrel{q\circ \pi}\to  C(\Xi)\rtimes_\alpha \ZM  \to 0
%\end{equation}
%which relates the topological invariants of the observable algebra for the quasiperiodic model $C(\Xi)\rtimes_\alpha\ZM$
%(e.g.\ Fibonacci) to those of $J$. The left vertical sequence implies that the topological invariants of $J$ come in parts from the edge algebra for the augmented version of the model $C(\tilde \Xi)\otimes\Kk$ and from the subalgebra $C_0(\Ss)\rtimes_\alpha\ZM$
%of the bulk algebra of the augmented version. The latter subalgebra
%can be seen as the algebra of perturbations we need to turn on to make a quasiperiodic into an augmented model.} 

%

\section{$K$-theory}
\label{sec-ktheory}

\subsection{Preliminaries}
\newcommand{\id}{\mbox{\rm id}}
\newcommand{\coker}{\mbox{\rm coker}}

We recall that any SES of $C^*$-algebras
$0\to I \stackrel{i}\to E \stackrel{q}\to A \to 0$
 gives rise to a $6$-term exact sequence in $K$-theory
$$\begin{array}{ccccc}
K_0(I) & \stackrel{i_*}\to & K_0(E) & \stackrel{q_*}\to & K_0(A) \\
\uparrow \mbox{\rm ind} & & & & \downarrow \exp \\
K_1(A) & \stackrel{q_*}\leftarrow  & K_1(E) & \stackrel{i_*}\leftarrow  & K_1(I)
\end{array}$$ 
This sequence yields a calculational tool to determine $K$-groups but we will see that in particular the exponential map has significant physical interpretation.

Applied to the Toeplitz extension (\ref{eq-TE}) the $6$-term exact sequence has the particular form (called Pimsner-Voiculescu exact sequence) \cite{PV}
$$\begin{array}{ccccc}
K_0(C(X)) & \stackrel{\id-\alpha_*}\to & K_0(C(X)) & \stackrel{\imath_*}\to & K_0(C(X)\rtimes_\alpha\ZM) \\
\uparrow  & & & & \downarrow \exp \\
K_1(C(X)\rtimes_\alpha\ZM) & \stackrel{\imath_*}\leftarrow  & K_1(C(X)) & \stackrel{\id-\alpha_*}\leftarrow  & K_1(C(X))
 \end{array}$$ 
It splits into two SESs
$$ 0\to C_{\alpha_*} K_i(C(X)) \stackrel{\imath_*}\to K_i(C(X)\rtimes_\alpha\ZM)\to I_{\alpha_*} K_{i-1}(C(X)) \to 0$$
for $i=0,1$. 
Here $C_\alpha$ is the coinvariant and $I_\alpha$ the invariant functor. More precisely, for a $\ZM$-modul $M$ with an isomorphism $\beta$, 
$$ I_\beta M = \ker (\id-\beta),\quad C_\beta M = \coker (\id-\beta) $$
%where $m\sim_\beta m'$ is the equivalence relation $\Rr\subset M\times M$ generated by 
%$(\beta(m),m)$ and 
\subsection{Application to Sturmian systems}
\newcommand{\nk}{\kappa}
Our interest is to analyse (\ref{eq-JE}) and here in particular its associated exponential map
$$ \exp_J : K_0(C(\Xi)\rtimes_\alpha\ZM) \to K_1(J)$$
(since there are more boundary maps around, we add for clarity the subscript $J$). For that 
we need first to determine $K_*(J)$ which can be done using the left vertical SES of (\ref{eq-dia}). We start therefore to investigate the $K$-theory of the ideal of that SES, that is, $K_*(C_0(\Ss)\rtimes_\alpha\ZM)$. $\Ss$ is parametrized by $(\phi,t)$ where $\phi$ is a singular value for the intercept and $t\in (0,1)$. Since the singular values correspond to the orbit $\theta\ZM$ of the rotation action on the circle, we may identify $\Ss = \ZM\times(0,1)$, and the $\ZM$ action on $\Ss$ is then simply given by the action of $\ZM$ on $\ZM$ in the left variable. It follows that $C_0(\Ss)\rtimes_\alpha\ZM\cong C_0(0,1)\otimes\Kk$ and thus
$$K_0(C_0(\Ss)\rtimes_\alpha\ZM) = 0,\quad K_1(C_0(\Ss)\rtimes_\alpha\ZM) \cong\ZM.$$
Moreover, the generator of $K_1(C_0(\Ss)\rtimes_\alpha\ZM)$ is given by the class of the function $f(n,t)= \delta_{n0} (e^{2\pi i t}-1)+1$ in the unitization of $C_0(\ZM\times(0,1))$. Furthermore, $q(f-1)=0$. Diagram (\ref{eq-dia}) shows therefore that $f-1$ lifts under $\pi$ to an element of $J$. It follows that the function $f$ lies in the kernel of the exponential map of the $6$-term exact sequence for the left vertical SES of (\ref{eq-dia}). In other words, the exponential map of the $6$-term exact sequence for the left vertical SES of (\ref{eq-dia}) is trivial. This implies, first, that the $6$-term exact sequence reduces to 
$$K_0(J) \cong K_0(C(\tilde\Xi)) \cong  \ZM$$
and
$$0\to K_1(C(\tilde\Xi)) \stackrel{\Ti_*}\to K_1(J) \stackrel{\pi_*} \to \ZM\to 0,$$
and second, that any element of $K_0(C(\Xi)\rtimes_\alpha\ZM)$ lifts to an element of 
$K_0(C(\tilde \Xi)\rtimes_\alpha\ZM)$ so that
the image of $\exp_J$ lies in the kernel of $\pi_*$. 
While the above sequence splits the isomorphism is not canonical. It depends on a choice of 
section $s:\ZM \to K_1(J)$, that is, preimage under $\pi_*$ of the generator of $\ZM$. Given such a section,  $\sigma:K_1(J)\to K_1(C(\tilde\Xi))$, $\sigma(x) =\Ti_*^{-1}( x - s\circ \pi_*(x))$ is a left inverse for $\Ti_*$ and
$$ K_1(J)\stackrel{(\sigma,\pi_*)}{\longrightarrow} K_1(\tilde\Xi)\oplus \ZM$$
an isomorphism.
The six term exact sequence of the SES (\ref{eq-JE}) can therefore be written
$$\begin{array}{ccccc}
K_0(C(\tilde\Xi)) & \stackrel{\Ti_*}\to & K_0(\Tt(\tilde\Xi,\alpha)) & \stackrel{q_*\circ\pi_*}\to & K_0(C(\Xi)\rtimes_\alpha\ZM) \\
\uparrow  & & & & \downarrow (\sigma,\pi_*)\circ \exp_J \\
K_1(C(\Xi)\rtimes_\alpha\ZM) & \stackrel{q_*\circ\pi_*}\leftarrow  & K_1(\Tt(\tilde\Xi,\alpha)) & \stackrel{\tilde\psi}\leftarrow & K_1(C(\tilde\Xi))\oplus \ZM
\end{array}$$
where $\tilde\psi([u]\oplus n) = \Ti_*([u]) + s(n)$. Under the isomorphism 
$K_i(\Tt(\tilde\Xi,\alpha)) \to K_i(C(\tilde\Xi))$ this becomes \cite{PV}
$$\begin{array}{ccccc}
K_0(C(\tilde\Xi)) & \stackrel{1-\alpha_*}\to & K_0(C(\tilde\Xi)) & \stackrel{q_*\circ\imath_*}\to & K_0(C(\Xi)\rtimes_\alpha\ZM) \\
\uparrow  & & & & \downarrow (\sigma,\pi_*)\circ \exp_J \\
K_1(C(\Xi)\rtimes_\alpha\ZM) &\stackrel{q_*\circ\imath_*}\leftarrow  & K_1(C(\tilde\Xi)) & \stackrel{\psi}\leftarrow & K_1(C(\tilde\Xi))\oplus \ZM
\end{array}$$
where $\psi([u]\oplus n) = (1-\alpha_*)([u]) + s(n)$. 
Since $\tilde \Xi$ is homeomorphic to the circle $\alpha_*$ is the identity on $K_i(C(\tilde\Xi))$, $i=1,2$.
Furthermore, $\mbox{\rm im} \exp_J = \ker \psi = K_1(C(\tilde\Xi))$ so that obtain 
\begin{equation}\label{eq-6S-cut}
\begin{array}{rcccccl}
0 \to & K_0(C(\tilde\Xi)) & \stackrel{q_*\circ i_*}\to & K_0(C(\Xi)\rtimes_\alpha\ZM) & \stackrel{\sigma\circ\exp_J}\to & K_1(C(\tilde\Xi)) &\to 0
\end{array}
\end{equation} 
The image of $q_*\circ i_*$ is generated by the class of the identity. The exponential map can be described as follows:
if a projection $p$ represents an element $[p]\in K_0(C(\Xi)\rtimes_\alpha\ZM)$, we lift $p$ to a self-adjoint element $a\in\Tt(\tilde\Xi,\alpha)$. Then $\exp_J([p])=[e^{2\pi ia}]\in K_1(J)$. By the above $\pi_*\circ \exp_J =0$ and hence there is even a representative $a$ such that $e^{2\pi ia}-1\in C(\tilde\Xi)\otimes \Kk$. Thus $\sigma\circ\exp_J([p])=[e^{2\pi ia}]\in K_1(C(\tilde\Xi))$.

\subsection{The exponential map for covariant families of operators}

We apply the exponential map to the $K_0$-class defined by the Fermi projection $P_F$
of a strongly pattern equivariant family of Hamiltonians $H_\Xi$ on $\ell^2(\ZM)$,
$$H_\xi = \left(\sum_{k\in S} (\widetilde {b_k})_\xi T^k\right) +  h.c.,\quad \xi\in \Xi$$
where $b_k$ are (finitely many) sliding block codes for $\Xi$. These represent an element in
$H\in C(\Xi)\rtimes_\alpha\ZM$. We suppose that the Fermi energy lies in a gap 
$\Delta=(E_0,E_1)$ of their (common) spectrum. Then also $P_F$ corresponds to an element in the crossed product and we wish to express its image under the exponential map in terms of  
a strongly pattern equivariant family of operators. We first note that
$$f(E) = \left\{\begin{array}{cl}
1 & \mbox{if } E\leq E_0 \\
\frac{E_1-E}{E_1-E_0} & \mbox{if } E\in (E_0,E_1) \\ 
0 & \mbox{if } E\geq E_1
\end{array} \right.$$
is a continuous function such that the Fermi projection of $H_\xi$ is given by
$$P_F(H_\xi) = f(H_\xi).$$
We can extend the sliding block codes $b_k$ to sliding block codes on $\tilde\Xi$ by linear interpolation: a word in a sequence of $\tilde\Xi$ may contain one single, or a pair of consecutive letters whose lengths differ from $a$ and $b$. These depend then on the parameter $t\in[0,1]$ and we can just extend $b_k$ on such a word by taking the convex combination of the extreme values. In this way we define the family $H_{\tilde\Xi}$.
 
Recall that the compression of the $H_{\tilde\xi}$ to $\ell^2(\NM)$,
\begin{equation}\label{eq-hs2}
\hat H_{\tilde\xi} = \Pi\left(\sum_{k\in S} (\widetilde {b_k})_{\tilde\xi} T^k\right)\Pi +  h.c.,\quad \tilde\xi\in \tilde\Xi
\end{equation}
are a strongly pattern equivariant family of Hamiltonians representing an element $\hat H$ of $\Tt(C(\tilde\Xi),\alpha)$. Furthermore, $\hat H$ is a lift of $H$ under the map $q\circ\pi$. We can therefore take $a = f(\hat H)$ and thus 
\begin{equation}
\label{eq-exp}
\sigma\circ\exp_J ([P_F(H_\Xi)]) = [e^{2\pi i f(\hat H_{\tilde \Xi})}] = 
%[\exp (2\pi i \frac{ E_1-\hat H_{\tilde\Xi} }{E_1-E_0}) P_\Delta(\hat H_{\tilde\Xi}) +  P_\Delta(\hat H_{\tilde\Xi})^\perp]
[e^{-i t_\Delta\hat H_{\tilde\Xi}} P_\Delta(\hat H_{\tilde\Xi}) +  P_\Delta(\hat H_{\tilde\Xi})^\perp]
\end{equation}
where $t_\Delta=\frac{ 2\pi i }{E_1-E_0}$ and we have used that $e^{i t_\Delta\hat E_1}$ is homotopic to $1$.  The boundary map applied to the $K_0$-class of the Fermi projection is thus the $K_1$-class of the unitary of time evolution defined by the family of half space operators $\hat H_{\tilde\Xi}$ by the charactistic time $t_\Delta$, 
restricted to the states in the gap. 

%%%%%%%%%%%%%%%%%%%%%

\section{Bulk boundary correspondance}
\label{sec-bulkboundary}
%%%%%%%%%%%%%%%%
\subsection{The winding number cocycle}\label{sec-8.1}
Recall that $\tilde\Xi$ is homeomorphic to the circle. It follows that $K_1(C(\tilde\Xi))$ is generated by the class of the function $S^1\ni z\mapsto z\in\CM^*$ once we have made an identification of  $\tilde\Xi$ with $S^1$. In particular, by taking the winding number of a unitary function representing the class of an element of $K_1(C(\tilde\Xi))$ one obtains a group isomorphism $\Ww: K_1(C(\tilde\Xi))\to\ZM$. 

Quite generally, a group homomorphism like $\Ww$ can be understood as a pairing of $K_1(C(\tilde\Xi))$ with a cyclice cocycle \cite{Connes}. In our case this cocycle 
$\eta$ is defined on the dense subalgebra 
$\Aa \subset A = C(\tilde\Xi)$ of continuous functions on $\tilde\Xi$ which are locally constant on the subspace $\Xi$, continuously differentiable on $\Ss=\tilde\Xi\backslash \Xi\cong \theta\ZM\times (0,1)$ and such that their derivative is integrable over $\Ss$. Here the integral over $\Ss$ is given by the Lebesgue integral on the components: $\int_\Ss f d\lambda = \sum_{n\in\ZM} \int_0^1 f(n\theta,t) dt$.
The winding number cocycle $\eta$ is now defined on the algebra $M_m(\Aa)$ of matrices with entries in $\Aa$ as
$$ \eta(f_1,f_2) = \frac1{2\pi i} \int_{\Ss}\Tr_m  f_1 \dot f_2 d\lambda $$   
where $\dot f_2$ is the derivative of $f_2$ w.r.t.\ $t$ and $\Tr_m$ the matrix trace on $m\times m$-matrices. In particular $\eta(f^{-1},f)$ is the winding number of an invertible function $f\in\Aa$ and
$$\Ww(y) =  \eta(f^{-1},f)$$
where $f$ is a representative in $M_m(\Aa)$ for $y$. 

We apply this to the unitary class given in (\ref{eq-exp}). 
The only spectrum of $\hat H_{\tilde\Xi}$ in the gap $\Delta$ is given by the Dirichlet eigenvalues $\mu_i$, $i=1,\cdots,m$. Hence, for singular $\phi$,
$$\Tr_m \big( e^{it_\Delta \hat H_{\phi,t}}\partial_t e^{-it_\Delta \hat H_{\phi,t}} P_\Delta( \hat H_{\tilde \Xi}) \big)= -it_\Delta \sum_{i=1}^m \partial_t {\mu_i}(\phi,t).$$
By Prop.~\ref{pro-KB} the set of values $\{\mu(x)|x\in \Xi\}$ has Lebegues measure $0$. The variation of $\mu(x)$ along $\Xi$ has thus zero weight and does not contribute to the winding number. If follows that 
%
%
%The restriction of $x\mapsto \mu(x)$ to $\Xi$ is a uniformly continuous function on a totally disconnected space and thus we can continuously deform it to a function which is locally constant on $\Xi$. In particular $[e^{it_\Delta \hat H_{\tilde\Xi}} P_\Delta( \hat H_{\tilde \Xi})]$ has a representative which belongs to $M_m(\Aa)$ and we can apply the winding number cocycle to this representative. 
%we obtain
%$$\Tr_m \big( e^{it_\Delta \hat H_{\phi,t}}\partial_t e^{-it_\Delta \hat H_{\phi,t}} P_\Delta( \hat H_{\tilde \Xi}) \big)= -it_\Delta \sum_{i=1}^m \partial_t {\mu_i}(\phi,t)$$
%and hence
$$\Ww(\sigma\circ\exp_J([P_F(H_\Xi)]) = \frac1{|\Delta|} \sum_{n\in\ZM}\int_0^1 \sum_{i=1}^m \partial_t {\mu_i}(n\theta,t) dt $$
This formula has a physical interpretation. Indeed, $\int_0^1 \partial_t \mu_i(n\theta,t) dt$ is the work performed on the $i$th boundary state in the gap $\Delta$ by the motion of moving an atom to perform the flip $ab$ to $ba$ at the position which is encoded by the singular value $\phi=n\theta$. $|\Delta|\Ww$ is therefore the total work performed by all such (phason) flips.
Note that 
the winding number can be inferred graphically from a plot like (Figure~\ref{Fig-EdgeSpectralFlow}) by reading off the spectral flow.
% one then has to take a homotopic representative of the class which lies in $M_m(\Aa)$.
\bigskip

It is not easy to compute analytically the winding number from the above explicit formula. 
We therefore make use of a very general result which allows to reformulate the winding number as a Chern number.
%Consider a $C^*$-dynamical system $(A,\alpha,\ZM)$. 

Dual to the boundary map $\delta:K_{i}(A\rtimes_\alpha\ZM)\to K_{i-1}(A)$ of the associated Pimsner-Voiculescu $6$-term exact sequence is a map $\#_\alpha:HC^i(A)\to  HC^{i+1}(A\rtimes_\alpha\ZM)$ on cyclic cohomology \cite{ENN,KRS}. In particular, if $\eta$ is a $1$-cycycle then this duality corresponds to 
%Applied to the situation in which $A = C(\tilde\Xi)$ and the winding number cocycle $\eta$ on $A$ one obtains 
the formula
\begin{equation}\label{eq-duality}
\eta(f^{-1},f) =  \#_\alpha\eta(p,p,p)
\end{equation}
provided $f$ is a representative for the $K_1$-class $\exp([p])$ defined by the projection $p\in A\rtimes_\alpha\ZM$.

Note that $\Aa$, the integral and the derivative are $\alpha$ invariant. The general theory therefore yields that $\#_\alpha \eta$ is the $2$-cocyle on $\Aa_\alpha\ZM$ given by \cite{KRS}
$$\#_\alpha \eta(a_1,a_2,a_3) = -i \Tr(a_1[\dot a_2,a_3'])$$
where the second derivation is given by $(f u^n)' = i n f u^n$ and the trace $\Tr$ by
$$\Tr(f u^n) = \delta_{n0} \int_\Ss f d\lambda .$$

\subsection{Gap-labelling}
%%%%%%%%%%%%%%%%
The dynamical system $(\Xi,\alpha)$ is uniquely ergodic. The unique ergodic probability measure $m$ defines a normalised trace $\tr:C(\Xi)\rtimes_\alpha\ZM\to\CM$ via
$$ \tr(\sum_n f_n u^n) = \int_\Xi f_0(\xi) dm(\xi).$$ 
Since traces are cyclic $\tr_*([p]) = \tr(p)$ is well-defined on classes of projections and so defines a tracial state $ \tr_*:K_0(C(\Xi)\rtimes_\alpha\ZM)\to\RM$.
The image of this tracial state is called the gap-labelling group. This name comes from the application to Schr\"odinger operators. Indeed, any spectral projection $P_{\leq E}$ of the operator onto its states of energy below $E$ is an element of the $C^*$-algebra, provided $E$ lies in the gap of the spectrum. The class of the spectral projection then defines an element in the $K_0$-group and $\tr_*([P_{\leq E}])$ equals the integrated density of states up to energy $E$ \cite{Bel95}.
  
For Sturmian systems the gap-labelling group is known to be
$\ZM+\theta\ZM$ and generated by its values on two projections, namely the identity $1\in C(\Xi)\subset  C(\Xi)\rtimes_\alpha\ZM$ and a projection $\chi_{[0^+,\theta^-]}\in C(\Xi)\subset  C(\Xi)\rtimes_\alpha\ZM$. More precisely, if we identify $\Xi$ with the cut up circle, then the indicator functions on subsets $[\phi_1^+,\phi_2^-]$ are continuous and hence projections, provided $ \phi_1$, and $\phi_2$ are singular values. The ergodic probability measure corresponds to the Lebesgue measure on the circle (normalised to $1$) and hence $\tr(\chi_{[\phi_1^+,\phi_2^-]}) = \phi_2-\phi_1$.

%%%%%%%%%%%%%%%%%%%%%%%%%%%%%%%%%%%%%%%%%%%%
%%%%%%%%%%%%%%%%%%%%%%%%%%%%%%%%%%%%%%%%%%%%

\subsection{The correspondance}
%%%%%%%%%%%%%%%%%%%%%%%%%%%%%%%%%%%%%%%%%
\begin{theorem}
%%%%%%%%%%%%%%%%%%%%%%%%%%%%%%%%%%%%%%%%
Let $p$ be a projection in $C(\Xi)\rtimes_\alpha\ZM$. There are $N,n\in\ZM$ such that
$$\tr_*([p]) = N+n\theta.$$
Moreover, $N$ is the unique integer such that $\tr_*([p]) \in ]0,1]$ and $-n$ is the winding number of $\sigma\circ\exp_J([p])$.
\end{theorem}
{\bf Proof}: 
We verify the result on the generators of $K_0(C(\Xi)\rtimes_\alpha\ZM)$.
We infer from (\ref{eq-6S-cut}) that there are two generators,  the generator coming from $K_0(C(\tilde\Xi))$ and the one from $K_1(C(\tilde\Xi))$. The first is given by
the class $[1]$ of the identity. It satisfies $\tr_*([1])=1$ and $\exp_J([1])=0$ and hence verifies the theorem.

Consider the projection $\chi_{[0^+,\theta^-]}$ whose trace is, as we have seen, $\theta$. We need to determine $\Ww(\sigma\circ\exp_J([\chi_{[0^+,\theta^-]}))$. For that we first show that 
$\chi_{[0^+,\theta^-]}$ lifts to a projection $P_\theta \in C(\tilde\Xi)\rtimes_\alpha\ZM$. 
Let $f$ be a real function on $\tilde\Xi$ which, in the open interval parametrized by $0\times (0,1)$ increases smoothly from $0$ to $1$, in the open interval parametrized by $\theta\times (0,1)$ decreases smoothly from $1$ to $0$, and is otherwise constant. In particular, $f$ 
restricts to $\chi_{[0^+,\theta^-]}$ on $\Xi$. 
We then set 
$P_\theta = gu + f + u^*g$ where 
$g=\sqrt{f(1-f)}\chi_{\theta\times (0,1)}$ (so the support of $g$ is where $f$ decreases to $0$). 
Clearly $q(g) = 0$ and $q(f) = \chi_{[0^+,\theta^-]}$ so $P_\theta$ is a lift of $\chi_{[0^+,\theta^-]}$. Moreover, $P_\theta$ is a projection and therefore $\sigma\circ \exp_J([\chi_{[\theta_i,\theta_i+\theta]}]) = \exp([P_\theta])\in K_1(C(\tilde\Xi))$ where $\exp$ is the exponential map for the Toeplitz extension of $(\tilde\Xi,\alpha)$. According to Lemma~\ref{lem-wind1} $\exp([P_\theta])$ has  winding number $-1$. It follows that $\exp([P_\theta])$ is a generator of $K_1(C(\tilde\Xi))$ and hence $\chi_{[0^+,\theta^-]}$ the other generator of $K_0(C(\Xi)\rtimes_\alpha\ZM)$. We have thus verified the statement of the theorem on the two generators.
\hfill q.e.d.
\bigskip

The reader may have noticed that the projection $P_\theta$ constructed in the proof of the last theorem is reminiscent of the Rieffel projection, the difference being that Rieffel's projection is an element of $C(S^1)\rtimes_\alpha\ZM$ where $\alpha$ is the rotation by $\theta$, a minimal action, whereas the action on $C(\tilde\Xi)$ is non-minimal.
\begin{lemma}\label{lem-wind1}
With the notation used in the proof of the last theorem $\exp[P_\theta]$ has winding number $-1$.  
\end{lemma}
{\bf Proof}: By (\ref{eq-duality}) the winding number of $\exp[P_\theta]$ is given by $-i\Tr(P_\theta[\dot P_\theta,P_\theta'])$
 where 
$\dot P_\theta = \dot gu + \dot f + u^*\dot g$ and $P_\theta' = i g u - i u^* g$. Then
\begin{eqnarray*}
-iP_\theta[\dot P_\theta,P_\theta'] &=& gu[-\dot f,u^*g] + f([u^*\dot g,gu] - [\dot g u, u^* g]) + u^*g [\dot f, gu] +R\\
& = & g^2(\dot f - \alpha(\dot f)) + \alpha^{-1}(g^2)(\alpha^{-1}(\dot f)-\dot f)
+ f(\alpha^{-1}(\dot {g^2}) - \dot {g^2}) + R
\end{eqnarray*}
where $R$ is a term which is proportional to non-zero powers of $u$ and hence vanishes under $\Tr$. 
%Using Davidson's notation $u g u^* = g_\theta$ we can rewrite this as
%$$-iP_\theta[\dot P_\theta,P_\theta'] -R = g^2(\dot f - \dot f_{\theta}) + g_{-\theta}^2(\dot f_{-\theta}-\dot f)+ f(\dot {g^2}_{-\theta} - \dot {g^2})$$
Since the integral is invariant under the action we can replace $\alpha^{-1}(g^2)(\alpha^{-1}(\dot f)-\dot f)$ by $g^2(\dot f - \alpha^{}(\dot f))$ and $f(\alpha^{-1}(\dot {g^2}) - \dot {g^2})$ by $(\alpha^{}(f) -f)\dot {g^2}=-(\alpha^{}(\dot f) -\dot f){g^2}+T$ where $T$ is a total derivative. 
%Up to a total derivative $(f_\theta -f)\dot {g^2}$ is given by $-(\dot f_\theta - \dot f){g^2}$. 
Thus
$$-i\Tr(P_\theta[\dot P_\theta,P_\theta']) = \int_\Ss 3 (\dot f - \alpha(\dot f)){g^2} d\lambda = 6 \int_0^1 \dot f(1,t){g^2(1,t)} dt$$
as $\alpha(\dot f) = -\dot f$ on the support of $g$. Finally
$\int_0^1 \dot f(1,t){g^2(1,t)} dt = \int_0^1 (-1)(t-t^2)dt = -\frac16$
and thus
$-i\Tr(P_\theta[\dot P_\theta,P_\theta']) = -1$. \hfill q.e.d.

\end{document}